\documentclass[
amsmath,amssymb,
aps,
prb,
reprint,
superscriptaddress,
]{revtex4-2}
\usepackage{amsmath, amssymb}
\usepackage{graphicx}
\usepackage{dcolumn}
\usepackage{bm}
\usepackage{mathtools}
\usepackage{hyperref}
\usepackage{xcolor}

\begin{document}

\title{Geometric theory of elastic curves constrained on rigid curved surfaces}

\author{Fahim Bin Selim}
\email{fahimbinselim@vt.edu}
\affiliation{Department of Physics, Virginia Polytechnic Institute and State University, Blacksburg, VA 24061, USA.}
\affiliation{Center for Soft Matter and Biological Physics, Virginia Polytechnic Institute and State University, Blacksburg, VA 24061, USA.}

\author{C. Nadir Kaplan}%
\email{nadirkaplan@vt.edu}
\affiliation{Department of Physics, Virginia Polytechnic Institute and State University, Blacksburg, VA 24061, USA.}
\affiliation{Center for Soft Matter and Biological Physics, Virginia Polytechnic Institute and State University, Blacksburg, VA 24061, USA.}
\affiliation{Department of Mechanical Engineering, Virginia Polytechnic Institute and State University, Blacksburg, VA 24061, USA.}
\affiliation{Center for the Mathematics of Biosystems, Virginia Polytechnic Institute and State University, Blacksburg, VA 24061, USA.}

\date{\today}

\begin{abstract}
Slender elastic objects constrained to curved surfaces are ubiquitous in soft systems, with examples ranging from DNA wrapping around histones to actin filaments forming contractile rings on the cell membrane during cytokinesis. In the continuum limit, the conformation of these effectively one-dimensional (1D) objects is governed by both the geometry and mechanics of the embedding surface, and the elasticity of the object. Existing phenomenological theories typically describe such behavior through elastic energy minimization that incorporates bending and twisting, but neglect stretching that may be important for filaments with finite cross-section or are highly incompatible with the underlying surface. Here we present a coarse-grained geometric theory of 1D elastic curves constrained on rigid surfaces, derived from the first principles. For an isotropic material, our model emerges naturally in terms of the Young’s modulus, Poisson’s ratio, and area moments of inertia. The resulting effective energy functional includes stretching in addition to bending and twisting. We minimize it to determine the equilibrium curve conformations on different rigid surfaces with zero, positive, or negative Gaussian curvature. Our theory provides a general framework for analyzing surface-bound elastic curves in biologically and physically relevant settings. 
\end{abstract}

\maketitle


\section{Introduction}
Filamentary objects are abound in biology: biopolymers, such as DNA, actin filaments, or microtubules are high-aspect-ratio elastic materials, which can stretch, bend, and twist~\cite{mehrbod2011significance,lazarus2015torsional, bicek2007analysis, yi2008buckling, shaevitz2008curvature, garrivier2000twisting, shi1994kirchhoff,  marko1998dna, marko1995statistical,  swigon1998elastic, dietz2009folding, enrique2010origin, goodsell1994bending}. Theoretically, these slender objects can be modeled as effectively 1D elastic curves, and their geometric degrees of freedom capture their mesoscopic behavior in space, where they assume a configuration free of mechanical stress in the absence of any geometric constraints~\cite{audoly2000elasticity}. Through different mechanisms, such as differential growth, chirality, or anisotropy, the filament may acquire intrinsic bending or twisting, mathematically captured by curvature and torsion functions $\bar{k}$ and $\bar{\tau}\,,$ respectively. Any deviation from the twisted and bent states would then cost energy~\cite{singer2008lectures, kamien2002geometry, marko1994bending, moroz1997torsional, yong2022statistics}.

\begin{figure}[b]
    \centering      \includegraphics[width=\columnwidth]{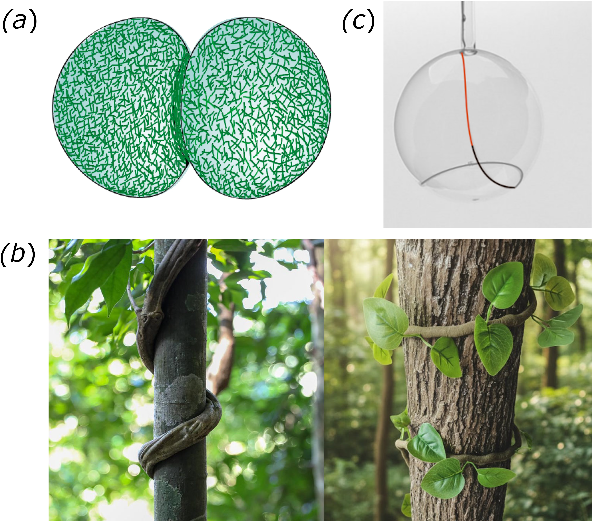}
    \caption{{\bf Slender biological and physical objects on curved surfaces.} \textbf{(a)} Formation of circular actomyosin contractile ring during cytokinesis on cell membrane~\cite{leite2019network}. \textbf{(b)} Climbing vines with a helical shape around tree trunks~\cite{liao2026vine}.  \textbf{(c)} A silicone-melt filament on soap bubble surface~\cite{prasath2021shapes}. (a), (b) are adapted, respectively, from the Refs. \cite{leite2019network}, \cite{liao2026vine} under the Creative Commons CC-BY 4.0 license. (c) is adapted from Ref.~\cite{prasath2021shapes} under the license from the Royal Society (Copyright 2021).} \label{fig:bio_examples}
\end{figure}

Rather than remaining immersed in the three-dimensional (3D) space, biological filaments are often confined to curved surfaces in many processes: DNA is wrapped in helical shape around histone octomer, actin filaments self assemble into a ring on the cell membrane to constrict the cell equator during cell division (cytokinesis) (Fig.~\ref{fig:bio_examples}a), or twinning plants climb support structures during growth (Fig.~\ref{fig:bio_examples}b)~\cite{bidone2014dynamic, dekraker2018regulation, miller2011contractile, leite2019network, gerbode2012cucumber, goriely2006mechanics, liao2026vine}. In these examples, the 3D conformation of the filament is dictated by the geometry of the constraining surface. The experimental setup in Fig.~\ref{fig:bio_examples}c illustrates this problem clearly~\cite{prasath2021shapes}: A flexible straight filament ($\bar{k}=0, \bar{\tau}=0$)  subject to gravity changes its shape when adsorbed to a spherical bubble of fixed radius. The surface constraint thus means that a space curve may not satisfy its intrinsic curvature and torsion while on the surface. Then, the curve is said to be ``incompatible" with the embedding surface, in turn acquiring residual stress at mechanical equilibrium due to the constrained degrees of freedom on the surface~\cite{ciarlet2005introduction}. 

\begin{figure}[t]
    \centering      \includegraphics[width=\columnwidth]{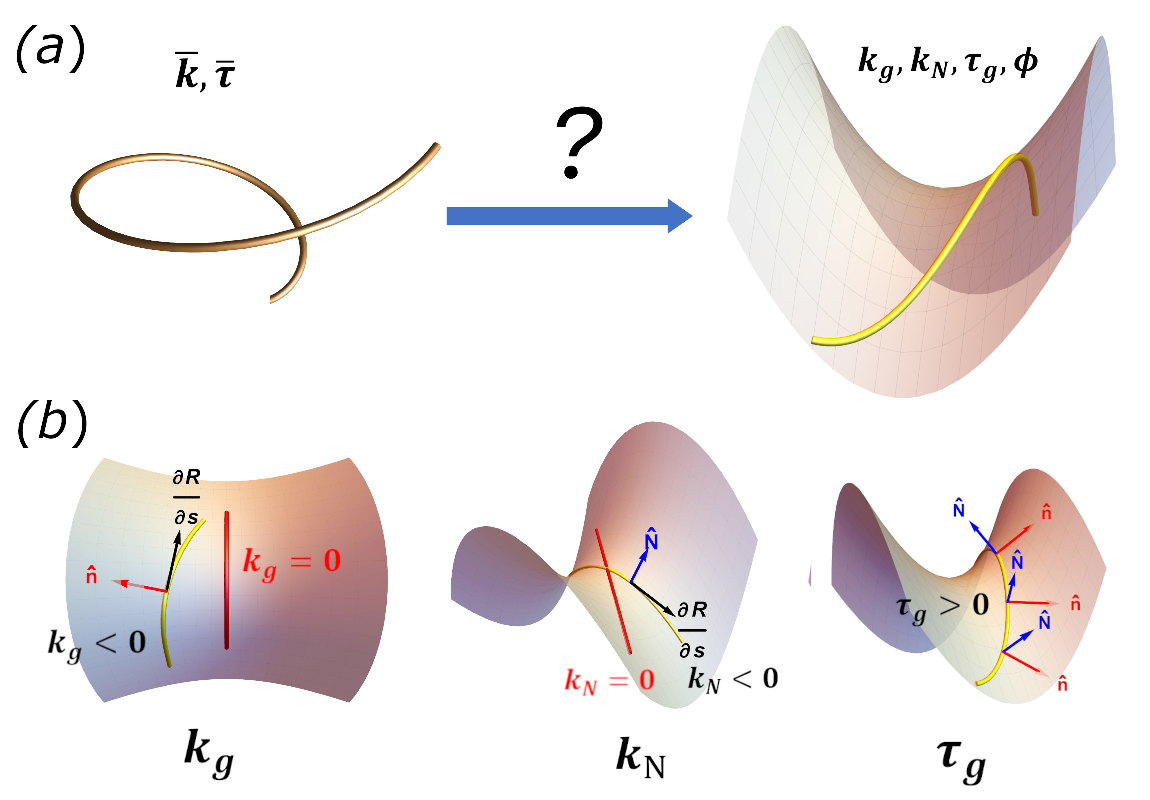}
    \caption{{\bf Curves in space versus on a surface.} 
    \textbf{(a)} Schematic illustration of the problem statement: An elastic space curve with given $\bar{k}$ and $\bar{\tau}$, if incompatible with a confining surface, undergoes elastic deformation to achieve an equilibrium surface configuration given by $k_g, k_N, \tau_g,$ and  the angle $\phi$ defined in Fig.~\ref{fig:DtoF}.\textbf{(b)} The geometric variables of a surface-bound curve: geodesic curvature $k_g$, normal curvature $k_N$, and geodesic torsion $\tau_g$.  \label{fig:problem_statement}}
\end{figure}

In this paper, we formulate a reduced hyperelastic theory to answer the following question motivated by the aforementioned biological examples: When an elastic space curve is embedded on a rigid surface constituting the ``reference" configuration, what is its preferred ``current" configuration at mechanical equilibrium (Fig.~\ref{fig:problem_statement}a)? Local alignment of the space curve (specified by two smooth functions $\bar{k}\,,\bar{\tau}$) tangent to the surface at a point determines the reference configuration as follows: $\bar{k}$ is expressed in terms of the ``in-plane" geodesic curvature $\bar{k}_g$ and the extrinsic normal curvature $\bar{k}_N$ coming from the local folding of the surface. Likewise, $\bar{\tau}$ can be given in terms of the geodesic torsion $\bar{\tau}_g$ and the rate of twist of the curvilinear coordinate system with a twist angle $\phi$ along a curve patch. The current configuration of the surface-bound curve at mechanical equilibrium, characterized by the analogous geometric variables $k_g, k_N, \tau_g$ and $\phi$ depends on the local features of the surface and the underlying constitutive hyperelastic model of choice (Fig.~\ref{fig:problem_statement}b). Importantly, the normal force acting on the curve -- to confine it to the surface -- breaks the directional symmetry across the cross-section of the filament represented by the curve, imposing hard restrictions on $k_N$ depending on the local mean curvature of the surface $H$. Similarly, the surface Gaussian curvature $K$ interrelates the variables $k_N$ and $\tau_g$ in the current configuration.

To tackle this problem, several studies previously employed a phenomenological elastic energy ($A_i:$ bending moduli as a function of the material elastic moduli and cross-sectional area moments of inertia of the filament, $ds:$ infinitesimal arc length of the curve) 
\begin{align}
    E = \int ds \left[ A_1 ( k_g - \bar{k}_g )^2 + A_2 ( k_N - \bar{k}_N )^2 + A_3 ( \tau_g - \bar{\tau}_g )^2 \right] \label{eq.2}
\end{align}
or similar formulations evaluated only on surfaces with both $H$ and $K$ constant, such as a sphere or cylinder~\cite{guven2014environmental, vazquez2015cylindrical, prasath2021shapes, sharma2023computational, wang2024deformation, wang2024equilibrium}. In all these cases, although the bending moduli $A_i$ must be proportional to the product of the Young's modulus $Y$ and the area moment of inertia $I$ of the slender material, the role of the Poisson's ratio of the material $\nu$ is not resolved, especially in $A_3\,.$ Furthermore, Eq.~\ref{eq.2} does not contain a stretching term that accounts for the local length change of the curve under longitudinal stress. This is because in unconstrained theories, the bending to stretching energy ratio of a filament of thickness $h$ and length $\ell$ scales as $(h/\ell)^2$~\cite{audoly2000elasticity}. When $h/\ell\ll1$, bending is the general preferential deformation, and stretching is much more costly, which still holds on surfaces with constant mean and Gaussian curvatures. However, under confinement, geometric frustration due to incompatibility alters the picture: if incompatibility with the bounding surface is strong for filaments with nonuniform intrinsic configuration, axial strain -- together with bending -- may relieve stress more than pure bending~\cite{marigo2006hierarchy}. 

To derive a hyperelastic theory applicable for a general curve confined on any curved smooth surface, we will reduce the elastic energy of 3D bodies first to an effective elastic theory for two-dimensional (2D) plates and shells through well-established coarse-graining procedures (see, e.g., Refs.~\cite{armon2011geometry, efrati2009buckling, efrati2009elastic, efrati2010non, efrati2013metric}), and then to 1D through a similar dimensional reduction. We will adopt a constitutive model constructed from the invariants of the Almansi strain where the metric tensor of the current configuration (that describes the local distances and angles of the material points) is used both in the elasticity tensor and as a volume measure~\cite{hanna2019some, wood2019contrasting}. Our goal is two-fold: First, we will incorporate local stretching of the curve inherited from the stretching terms in the 2D theory, second, derivation of the theory from first principles will naturally express the stretching and bending moduli in terms of the dimensions of an elastic filament, as well as the Young's modulus and the Poisson's ratio of the filament material. Throughout this roadmap and when considering surface-bound elastic curves, we will use differential geometry to describe curves, surfaces, and curves on surfaces~\cite{ciarlet2005introduction, stoker2011differential, siegel2008gaussian, kaplan2013intrinsic}.


This paper is organized as follows: In Sec.~\ref{sec.2}, we introduce 3D and 2D Riemannian manifolds, as well as curves in space and bounded on surfaces. In Sec.~\ref{sec.3} and~\ref{sec.4}, we first derive an effective theory of incompatible elasticity for slender objects modeled as curves and then derive a theory for incompatible curves bounded on curved surfaces. We demonstrate the power and generality of our theory with numerical results for curves with different intrinsic curvature and torsion pairs bounded on a wide range of surfaces in Sec.~\ref{sec.5}. In Sec.~\ref{sec.6}, we discuss the results and potential future directions for this work.

\section{Geometry of elastic curve and embedding surface} \label{sec.2}

To derive a theory for 1D elastic space curves confined to 2D Riemannian manifolds, we will first define the geometric variables that specify a reference curve (i.e., the space curve in the reference configuration) in the Frenet-Serret frame embedded in a 3D Euclidean space. Then, we will define a 3D Riemannian manifold  for a bulk material, which will be used to coarse grain a given 3D constitutive model into 1D to assign the curve its elastic properties. After that, we will introduce a 2D Riemannian manifold that will serve as the rigid embedding surface of the elastic curve. Finally, we will define the Darboux frame, which is the most natural representation of the curve in its surface-bound state. We will formulate the local embedding through a coordinate transformation from the Frenet-Serret apparatus for the space curve to the Darboux frame for its surface-bound configuration.

Throughout this work, we will employ the following conventions: Latin letters will be used as indices for the components of objects (vectors and tensors) defined in 3D, e.g., $i, j, k\in\{1, 2, 3\}$, and Greek letters for the components of objects defined in 2D, e.g., $\alpha, \beta, \gamma\in\{1, 2\}.$ Following Einstein convention, repeated indices imply summation. For any rank-2 tensor $\mathbf{A}\in \mathbb{R}^3\otimes\mathbb{R}^3$  or $\mathbf{B}\in \mathbb{R}^2\otimes\mathbb{R}^2$, the determinants are labeled as $|A|=\det(\mathbf{A})$, $|B|=\det(\mathbf{B})\,.$ For partial derivatives in 1D, we will use the short-hand notation $\partial_x\equiv \partial/\partial x$ where $x$ is the unitless or arc length parametrization along the curve. In 2D or 3D, we will use the short-hand notation $\partial_i\equiv \partial/\partial x^i$ or $\partial_\alpha\equiv \partial/\partial x^\alpha\,,$ where $x^i\,,x^\alpha$ stand for the curvilinear coordinates in a given 2D or 3D manifold. Higher order derivatives will then be denoted by, e.g., $\partial_{i_1\dots i_n}\equiv\partial/\partial x^{i_1}\dots\partial x^{i_n}\,,$ where $\{x^{i_1},\dots x^{i_n}\}$ is a set of coordinates. The $\nabla$ operator will represent the covariant derivatives on the Riemannian manifolds. For the reference curve, we will indicate the corresponding variables by using an overline, e.g., $\bar{s}$ for the curvilinear arc length coordinate along the curve.




For a reference curve with the position vector $\mathbf{\bar{R}}(\sigma)\in\mathbb{R}^3$, the metric is defined as $\bar{g}\equiv\partial_{\sigma}\mathbf{\bar{R}} \cdot \partial_{\sigma}\mathbf{\bar{R}}$ ($\sigma \in[0,1]:$ fixed coordinate along the curve). An arc length coordinate $\bar{s}$ can be defined such that the unit tangent vector to the curve is given by 
\begin{equation}
\partial_{\bar{s}}\mathbf{\bar{R}}=\frac{1}{\sqrt{\bar{g}}}\partial_{\sigma}\mathbf{\bar{R}}. 
\label{eq:ref_arc_length}
\end{equation}
In the arc length representation, the curve normal and binormal vectors are respectively defined as $\mathbf{\bar{L}}\equiv\frac{\partial_{\bar{s}\bar{s}}\mathbf{\bar{R}}}{||\partial_{\bar{s}\bar{s}}\mathbf{\bar{R}}||}$ and $\mathbf{\bar{B}}\equiv\partial_{\bar{s}}\mathbf{\bar{R}} \times \mathbf{\bar{L}}$, constituting the orthonormal triad of the Frenet-Serret vectors ($\partial_{\bar{s}}\mathbf{\bar{R}}\perp \; \mathbf{\bar{L}} \perp \;\mathbf{\bar{B}}$), setting the material frame at each point of the space curve \cite{kamien2002geometry}. Then, the definitions of the curvature and torsion of the space curve follow as $\bar{k} \equiv \partial_{\bar{s}\bar{s}}\mathbf{\bar{R}} \cdot \mathbf{\bar{L}}$, $\bar{\tau} \equiv \partial_{\bar{s}}\mathbf{\bar{L}} \cdot \mathbf{\bar{B}}$ (Appendix~\ref{ap B}). Given any three smooth functions $\bar{g}, \bar{k}, \bar{\tau}$, a unique curve in space can be constructed up to rigid translations and rotations, ensuring curve compatibility in $\mathbb{R}^3$~\cite{stoker2011differential}. 



To construct a general theory of surface-bound elastic curves, we next introduce the manifold of a 3D elastic body that will be coarse-grained into a quasi-1D filament: Let a 3D domain $\mathcal{D} \subset \mathbb{R}^3$, with immersion in the ambient 3D Euclidean space, be represented with the position vector $\mathbf{Y}:\mathcal{D} \rightarrow \mathbb{R}^3$. If $\mathcal{D}$ has the local material curvilinear coordinates $\mathbf{x}=(x^1, x^2,x^3),$ the manifold has tangent vectors $\partial_i\mathbf{Y}$. Then, we can define the metric $g_{ij} \equiv \partial_i \mathbf{Y} \cdot \partial_j \mathbf{Y}$, which is a $3\times 3$ symmetric, positive-definite tensor that makes $\mathcal{D}$ a Riemannian manifold; the inverse metric tensor $g^{ij}$ such that $g^{ij}g_{jk}=\delta^i_k$; and the affine connection tensor $\Gamma^i_{kl} \equiv g^{im} \partial_{kl} \mathbf{Y} \cdot \partial_m \mathbf{Y}$ known as the Christoffel symbols. Furthermore, we define the covariant derivative as $\nabla_i A^{jk} \equiv \frac{1}{\sqrt{|g|}} \partial_i(\sqrt{|g|}A^{jk}) - \Gamma^j_{il}A^{kl}$ for a rank-2 tensor $\mathbf{A}\,.$ Appendix~\ref{ap D} gives the 3D elastic energy that utilizes $g_{ij}, g^{ij}$, and $\nabla_i\,,$ along with the corresponding Euler-Lagrange equations that yield the equilibrium configuration of a bulk material.


For the rigid embedding surface of the elastic curve, a 2D domain $\mathcal{S} \subset \mathbb{R}^3$ is given by the map $\mathbf{X}(x^1, x^2):\mathcal{S}\rightarrow \mathbb{R}^3$ with the surface coordinates $x^1,\,x^2.$ Then, we can define two tangent vectors $\partial_{\alpha}\mathbf{X}$ and the 2D metric tensor $a_{\alpha\beta}\equiv\partial_{\alpha}\mathbf{X} \cdot \partial_{\beta}\mathbf{X}\,.$ The normal vector at each point on the surface is defined by $\mathbf{\mathbf{\hat{N}}} \equiv \left(\partial_1\mathbf{X} \times \partial_2\mathbf{X}\right)/\sqrt{|a|}$. In 2D, the Christoffel symbols $\Gamma^{\alpha}_{\beta\gamma}$ and the second fundamental form $b_{\alpha\beta}$ are specified by the relation~\cite{stoker2011differential}
\begin{equation}
\partial_{\alpha \beta}\mathbf{X} \equiv \Gamma^{\gamma}_{\alpha \beta} \partial_{\gamma}\mathbf{X} + b_{\alpha \beta} \mathbf{\hat{N}}\,.
\label{eq:2nd_deriv}
\end{equation}
In what follows, we will use these definitions not only for the embedding surface, but also when reducing the 3D constitutive model to an effective 2D theory for elastic shells, which will be coarse-grained once more to formulate a first-principles theory of slender elastic filaments.

\begin{figure}[b]
    \centering      \includegraphics[width=\columnwidth]{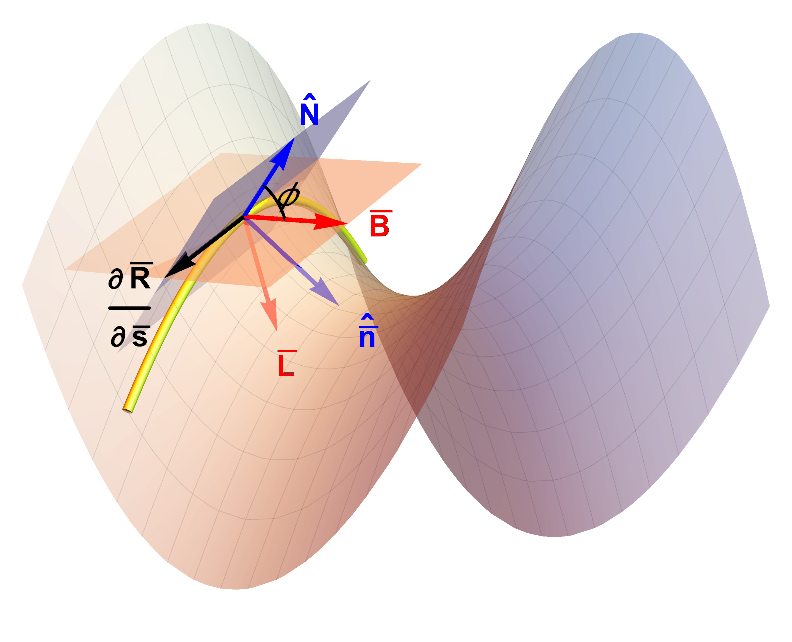}
    \caption{\small {\bf Frenet-Serret versus Darboux frames along a space curve.} The rotation angle $\phi$ between the the Darboux frame (blue arrows) of the curve on the surface (yellow) and the Frenet-Serret frame (red arrows) of the same curve -- if viewed without the surface in space. The curve tangent vector $\frac{\partial \mathbf{R}}{\partial \bar{s}}$(black) is common to both frames and simultaneously orthonormal to all the other vectors.}
    \label{fig:DtoF}
\end{figure}

When the reference curve is constrained at a given point on the surface $(x_0^1\,, x_0^2)$ with the normal vector $\mathbf{\hat{N}}\,,$ such that $\mathbf{\bar{R}}(\sigma_0)\equiv\mathbf{X}(x_0^1(\sigma_0)\,, x_0^2(\sigma_0))\,,$ the natural definition of the curve normal tangent to the surface  $\mathbf{\hat{n}}$ leads to $\mathbf{\hat{\bar{n}}} \equiv \mathbf{\hat{N}} \times \partial_{\bar{s}} \mathbf{\bar{R}}\,.$ This intuitive choice defines the Darboux frame where the curve tangent $\partial_{\bar{s}} \mathbf{\bar{R}}$, the ``second" tangent $\mathbf{\hat{\bar{n}}}\,,$ and the surface normal $\mathbf{\hat{N}}$ form an orthonormal triad, i.e.,  $\partial_{\bar{s}}\mathbf{\bar{R}}\perp \; \mathbf{\hat{\bar{n}}} \perp \;\mathbf{\hat{N}}$). The Darboux frame varies from the Frenet-Serret frame by a twist angle $\phi$ about the curve tangent, such that (Fig. \ref{fig:DtoF}) \cite{guven2014environmental, vazquez2015cylindrical},
\\
\begin{equation}
\begin{bmatrix}
\mathbf{\partial_{\bar{s}}\bar{R}} \\  \\
\mathbf{\hat{\bar{n}}} \\ \\
\mathbf{\hat{N}} 
\end{bmatrix} 
=
\begin{bmatrix}
1 && 0 && 0 \\ \\
0 && \cos\phi && -\sin\phi \\ \\
0 && \sin\phi && \cos\phi
\end{bmatrix}
\begin{bmatrix}
\mathbf{\partial_{\bar{s}}\bar{R}} \\  \\
\mathbf{\bar{L}} \\ \\
\mathbf{\bar{B}} 
\end{bmatrix} \,.
\label{eq:coordtransform}
\end{equation}

As with the definitions of the curvature $\bar{\kappa}$ and torsion $\bar{\tau}$ in the Frenet frame, similar scalar geometric variables can be introduced through the higher order derivatives of the Darboux frame vectors at a given point on the curve. These variables are the geodesic curvature $\bar{k}_g\equiv \partial_{\bar{s}\bar{s}}\mathbf{\bar{R}}\cdot\mathbf{\hat{\bar{n}}}$, normal curvature $\bar{k}_N\equiv\partial_{\bar{s}\bar{s}}\mathbf{\bar{R}}\cdot\mathbf{\hat{N}}$, and geodesic torsion $\bar{\tau}_g\equiv\partial_{\bar{s}}\mathbf{\hat{\bar{n}}}\cdot\mathbf{\hat{N}}$ for the surface-bound reference curve (Fig.~\ref{fig:problem_statement}b). Then, combining the Darboux equations for the reference curve on the surface
    \begin{equation}
        \begin{bmatrix}
    \partial_{\bar{s}\bar{s}} \mathbf{\bar{R}} \\  \\
    \partial_{\bar{s}} \mathbf{\hat{\bar{n}}} \\ \\
    \partial_{\bar{s}} \mathbf{\hat{N}} 
    \end{bmatrix} 
    =
    \begin{bmatrix}
    0 && \bar{k}_g && \bar{k}_N \\ \\
    -\bar{k}_g && 0 && \bar{\tau}_g \\ \\
    -\bar{k}_N && -\bar{\tau}_g && 0
    \end{bmatrix}
    \begin{bmatrix}
    \partial_{\bar{s}} \mathbf{\bar{R}} \\  \\
    \mathbf{\hat{\bar{n}}} \\ \\
    \mathbf{\hat{N}} 
    \end{bmatrix} 
    \label{eq:Darboux}
    \end{equation}
    with Eq.~\ref{eq:coordtransform} yields the relations
    \begin{gather}
        \bar{k}_g = \bar{k} \cos\phi, \label{eq.27}\\
        \bar{k}_N = \bar{k} \sin\phi, \label{eq.28}\\
        \bar{\tau}_g = \bar{\tau} - \frac{\partial \phi}{\partial \bar{s}}, \label{eq.29} 
    \end{gather}
    so that $\bar{k}^2=\bar{k}^2_g+\bar{k}^2_N\,.$ Note that, while the analytic functions $\bar{g}\,,\bar{k}\,,\bar{\tau}$ describe a reference curve in space, the variables representing the same curve on the surface $\bar{g}\,,\bar{k}_g\,,\bar{k}_N\,,\bar{\tau}_g$ are not necessarily smooth since $\phi$ may become nonanalytic when finding the mapping in Eqs.~\ref{eq.27}--\ref{eq.29} throughout the curve. This means that, under surface compatibility conditions that demand $C^3$ continuity of $\mathbf{\bar{R}}(\sigma)$ along the coordinate patch $(x^1(\sigma),x^2(\sigma)),$ the reference space curve may not be ``immersible" on the surface. This is what will give the elastic curve its residual stress when finding its current configuration corresponding to $g, k_g, k_N, \tau_g$ on the surface at equilibrium, which must be immersible.

\section{Reduced Theory of Incompatible 1D Elasticity}
\label{sec.3}

\begin{figure*}[ht]
        \centering      \includegraphics[width = \textwidth]{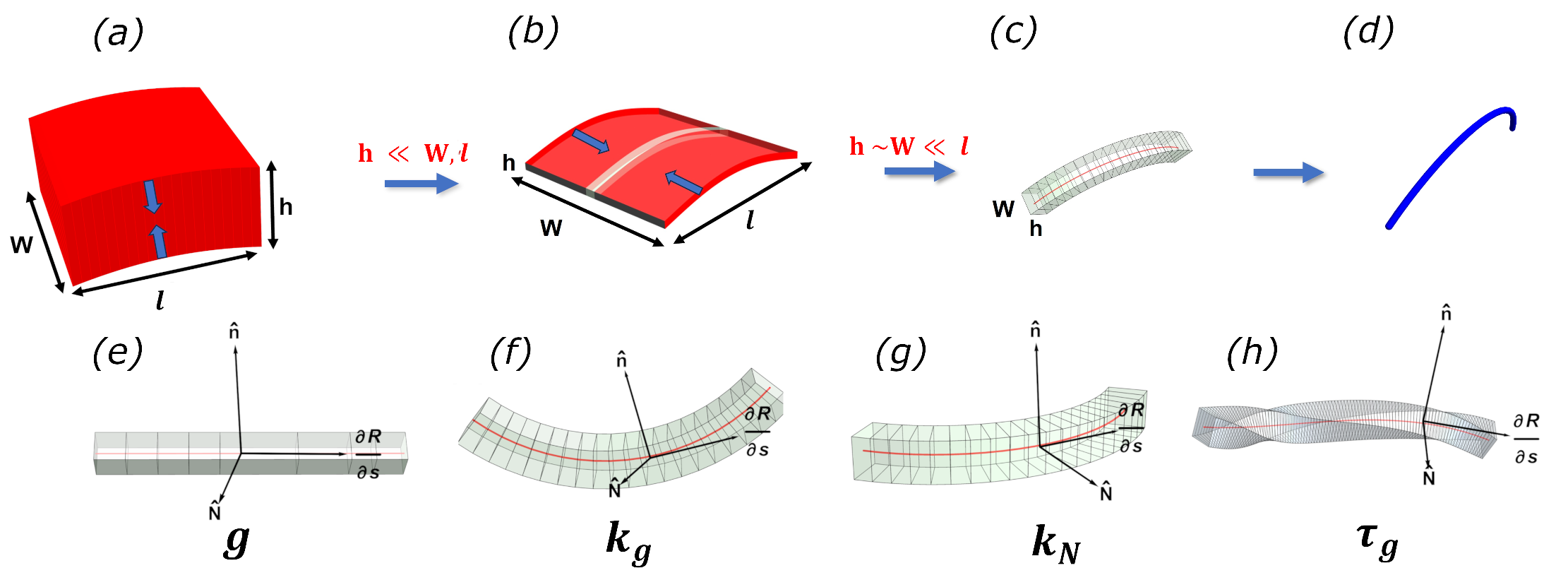}
        \caption{\textbf{Coarse-graining a 3D elastic body to a 1D filament.} Coarse-graining sequence from \textbf{(a)} 3D first to \textbf{(b)} 2D and then to \textbf{(c), (d)} 1D. The dimensions are defined as $\ell:$ length,  $h:$ thickness, and $W:$ width. The elastic behavior of the slender rod in (c) can be effectively captured by the deformations of its midcurve (red), and the integration over the thin dimensions leads to a coarse-grained curve (blue) in (d). (e)--(h) Admissible deformations in the elastic energy density (Eq.~\ref{eq.23}): \textbf{(e)} axial stretch $g$, \textbf{(f)} geodesic curvature $k_g$, \textbf{(g)} normal curvature $k_N$, and \textbf{(h)} geodesic torsion $\tau_g$.}
        \label{fig:coarse}
    \end{figure*}

    For a rigorous derivation of a 1D elastic energy whose minimum will give the current configuration of the curve, we begin from an incompatible, covariant formulation in 3D that measures the energy cost with respect to the generally incompatible reference configuration~\cite{ciarlet2005introduction}. Previously, by small thickness approximation and dimensional reduction, a 2D effective theory was derived and applied to elastic plates and shells that acquire residual stress due to their incompatibility~\cite{efrati2009elastic, efrati2010non, efrati2009buckling, efrati2013metric, armon2011geometry, grossman2016elasticity}. Even though this theory -- starting from an energy constituted from the invariants of the Green strain that yields a rank-4 elasticity tensor written in terms of the reference metric $\bar{g}_{ij}$ -- is very successful in describing the physical behavior of 3D elastic materials, difficulties arise when reducing it to an effective theory for lower-dimensional elastic objects, such as rods~\cite{hanna2019some}. The choice of the elasticity tensor at the heart of this constitutive model leads to the following physical outcomes in 1D: (a) there will be no pure curvature terms in the energy such that the change of the rod length costs not only stretching, but also bending energy; (b) the rod can be compressed to zero length with finite energy cost (Appendix~\ref{ap E}). Therefore,  starting from an energy composed of the invariants of the Almansi strain by following Hanna et al., we use the current metric $g_{ij}$ both as the volume measure and in the rank-4 elasticity tensor $A^{ijkl}$~\cite{hanna2019some, wood2019contrasting}. Then, in the material curvilinear coordinates $\mathbf{x}=(x^1, x^2, x^3)$, and with Lam\'e constants $\lambda$ and $\mu$, the differential volume element, the elasticity tensor in 3D, and the 3D strain tensor are
    \begin{gather}
        dV = \sqrt{|g|} \; dx^i \nonumber\,,  \\
        A^{ijkl} \equiv  \left( \lambda \; g^{ij} g^{kl} + \mu \; \left( g^{ik} g^{jl} + g^{il} g^{jk} \right) \right)\,, \label{eq:elasticity_tensor_main}  \\ 
        \epsilon_{ij} \equiv \frac{1}{2} (g_{ij} - \bar{g}_{ij}).\label{eq:strain_tensor_main}  
    \end{gather}  
    These quantities allow us to write the 3D elastic energy functional,
    \begin{gather}    
        E_{3D} = \int dV \frac{1}{2} \sqrt{\frac{|\bar{g}|}{|g|}} A^{ijkl} \epsilon_{ij} \epsilon_{kl}\,, \nonumber \\   
        = \int \left( \sqrt{|\bar{g}|} \prod_{i} dx^i \right) \frac{1}{2}  A^{ijkl} \epsilon_{ij} \epsilon_{kl}\,,  \nonumber \\      
        \Rightarrow E_{3D} = \int d\bar{V} \frac{1}{2}  A^{ijkl} \epsilon_{ij} \epsilon_{kl}\,. \label{eq.3}
    \end{gather}
    To derive a 2D theory for elastic shells, we integrate Eq.~\ref{eq.3} over the height $h$ of an elastic shell, which is much smaller than its width $W$ and length $\ell$ ($z\equiv x^3 \in [-h/2, h/2]$ and $h \ll W,\ell\,;$ Fig.~\ref{fig:coarse}a,~b). In the thin shell limit, a plausible approximation is that the natural boundary conditions effectively hold across the height of the 3D object. This approximation imposes the two Kirchoff-Love assumptions for plane stress and plane strain~\cite{efrati2009elastic}. The plane stress assumption enforces the stress to remain parallel to the deformed midsurface of the shell. In the plain strain assumption, no longitudinal shear deformations may occur across the midsurface. Under these two assumptions, integrating out the thin dimension in Eq.~\ref{eq.3} about the midsurface at $z=0$ yields the reduced 2D energy functional in terms of the stretching content $E_s$ and the bending content $E_b$ ($|\bar{a}|\,,|a|$ determinants of the metric tensors of the midsurface in the reference and current configurations),
    \begin{equation}
    \begin{split}
        E_{2D} = & \int dA~\sqrt{\frac{|\bar{a}|}{|a|}}   w_{2D}, \\
        \omega_{2D} & \equiv \frac{h}{2} E_s + \frac{h^3}{24} E_b\,.
    \end{split}
    \label{eq.4}
    \end{equation}
    By defining the 2D elasticity tensor $\mathcal{A}^{\alpha \beta \gamma \delta}$ and the 2D strain tensor $\varepsilon_{\alpha \beta}$ ($a_{\alpha \beta}, b_{\alpha \beta}:$ the metric and second  fundamental form of the midsurface in the current configuration, $\bar{a}_{\alpha \beta}, \bar{b}_{\alpha \beta}:$ their counterparts in the reference configuration, $Y:$ Young's modulus, $\nu:$ Poisson's ratio)
     \begin{gather}
        \mathcal{A}^{\alpha \beta \gamma \delta} \equiv \frac{Y}{8(1-\nu^2)} \left( \nu \; a^{\alpha \beta} \; a^{\gamma \delta} + (1-\nu) \; a^{\alpha \gamma} \; a^{\beta \delta} \right), \\
        \varepsilon_{\alpha \beta} \equiv \frac{1}{2}(a_{\alpha \beta}-\bar{a}_{\alpha \beta}),
        \label{eq:elasticity_tensor}
    \end{gather} 
    the stretching content $E_s$ and the bending content $E_b$ under the small stretch approximation $\delta\equiv a^{11}\varepsilon_{11} \ll 1$ are found as ($b^2:$ quadratic terms in the elements of $b_{\alpha\beta}$)
    \begin{equation}
        \begin{split}
    E_s & = \mathcal{A}^{\alpha \beta \gamma \delta}
    \varepsilon_{\alpha \beta}\;\varepsilon_{\gamma \delta}\,,\\
    E_b & = \mathcal{A}^{\alpha \beta \gamma \delta}
    (b_{\alpha \beta}-\bar{b}_{\alpha \beta})\;(b_{\gamma \delta}-\bar{b}_{\gamma \delta}) + \mathcal{O}(b^2\delta^2)\,.  
    \end{split}
    \label{eq.5}
    \end{equation}
    The derivation of Eqs.~\ref{eq.4}--\ref{eq.5} is detailed in Appendix~\ref{ap D}.
    
   

    Starting from the reduced 2D energy functional given by Eq.~\ref{eq.4}, we can derive an effective 1D theory for filamentary objects with two of their dimensions much smaller than their length ($h,W \ll \ell$, Fig. \ref{fig:coarse}b--d) by following the same blueprint of Appendix~\ref{ap D}. When the elastic behavior of a thin 2D body is effectively described by its midsurface with orthogonal coordinates $ (\sigma, t)\,,$ we can take a small strip from the midsurface (i.e., $W\equiv\Delta t\ll \ell$). and approximate the elastic behavior of the strip with the elastic behavior of its midcurve positioned at $t=0$  (Fig.~\ref{fig:coarse}b,~c). This positions the long axis of the strip along the coordinate $\sigma$, and $t$ points in the direction perpendicular to the strip. 
    
    Coarse-graining from the 2D shell to the 1D strip (shown in Fig.~\ref{fig:coarse}d in blue) requires the 1D equivalent of the Kirchoff-Love assumptions for the stress and moments, which we can introduce by using the natural boundary conditions of the 2D theory across the width of the filamentary object. To find these natural boundary conditions, the Euler-Lagrange equations for the 2D moment $\mathcal{L}^{\mu \nu}$ and 2D augmented stress $\mathcal{K}^{\mu \nu}$
    \begin{gather}
     \mathcal{L}^{\mu \nu} = \sqrt{\frac{|\bar{a}|}{|a|}} \frac{\partial w_{2D}}{\partial b_{\mu\nu}}\,,\\   
     \mathcal{K}^{\mu \nu} = \sqrt{\frac{|\bar{a}|}{|a|}} \left( \frac{\partial w_{2D}}{\partial \varepsilon_{\mu\nu}} + \frac{\partial w_{2D}}{\partial a_{\mu\nu}} \right)\,,
    \end{gather}   
    can be derived by extremizing Eq.~\ref{eq.4}. This leads to Eqs.~\ref{D.11} and~\ref{D.12} (Appendix~\ref{ap D}) along with the natural boundary conditions Eqs.~\ref{D.13}--\ref{D.15}, which are the surface terms arising from the variation of Eq.~\ref{eq.4}. To derive the 1D analog of Kirchoff-Love assumptions, we will use two of these natural boundary conditions in the algebraic form ($n_\alpha:$ components of the curve normal tangent to the surface)
    \begin{gather}
    \left( \mathcal{K}^{\mu\nu} +\mathcal{L}^{\gamma \nu}b_{\gamma}^{\mu} \right)  n_{\nu} = 0\,,
      \label{eq.7}\\
         \mathcal{L}^{\mu \nu} n_{\mu} n_{\nu} = 0\,. \label{eq.6}
    \end{gather}   
    Here, with the current metric as the measure,  $b^{\mu}_{\gamma} = a^{\alpha \mu}b_{\alpha \gamma}$. As with the 2D Kirchoff-Love assumptions, we assume that Eqs.~\ref{eq.7},~\ref{eq.6} hold across the strip, such that $n_1=0$ and $n_2=1$ (i.e., the normal vector to the strip is aligned everywhere along the $t-$coordinate). Then, Eq.~\ref{eq.7} becomes
    \begin{equation}
        \begin{split}
        &\left( \mathcal{K}^{\mu 2} +\mathcal{L}^{\gamma 2}b_{\gamma}^{\mu} \right)  n_{2} = 0\,,\\
        &\Rightarrow \mathcal{K}^{\mu 2} +\mathcal{L}^{\gamma 2} b_{\gamma}^\mu = 0\,, 
    \end{split}
    \label{eq.8a}
    \end{equation}  
    which gives two approximations for the components of the strain tensor:

    \noindent
    \textit{(a)\textit{ Zero shear strain:}} 
    For $\mu=1$ in Eq.~\ref{eq.8a},
        \begin{gather}
        \mathcal{K}^{12} +\mathcal{L}^{12}a^{11}b_{11} = 0\,, \nonumber \\
        \Rightarrow \varepsilon_{12} = 0+ \mathcal{O}(h^2)\,, \label{eq.8}
    \end{gather}   
    implying that the shear strain for the slender object can be neglected.

    \noindent
    \textit{(b) Modified linear strain :}   
    For $\mu=2$ in Eq.~\ref{eq.8a},
    \begin{gather}
        \mathcal{K}^{22} + \mathcal{L}^{12}b_1^2 +\mathcal{L}^{22}b_2^2 = 0, \nonumber \\
        \Rightarrow a^{22}\varepsilon_{22} = -\nu a^{11}\varepsilon_{11} + \mathcal{O}(\gamma^2, \delta^2), \label{eq.9} 
    \end{gather}
    which relates the transverse strain to the axial strain through the Poisson's ratio $\nu$, giving a modified linear strain relation. Here, we impose the approximation -- in addition to the small thickness and small width approximations $\frac{h}{\ell} = \frac{W}{\ell} \equiv \gamma \ll 1$ -- that the axial stretch is small ($ \delta \ll 1$), as we shall see that the stretching energy density of the strip that scales with $YhW$ is much bigger than the bending energy density that scales with $Y h^3W/\ell^2\,,$ rendering higher order contributions negligible.

    \noindent
    \textit{(c) Zero transverse bending moment:} Eq.~\ref{eq.6} becomes
    \begin{equation}
        \mathcal{L}^{22} n_{2} n_{2} = \mathcal{L}^{22}=0\,,    
    \end{equation}
    which gives,
    \begin{equation}
    a^{22}(b_{22}-\bar{b}_{22}) = - \nu a^{11}(b_{11}-\bar{b}_{11})\,, \label{eq.10}
    \end{equation}
    expressing the bending deformation in the transverse direction in terms of the one in the axial direction. Hence, our theory accounts for finite transverse strain and bending moments through Eqs.~\ref{eq.9}--\ref{eq.10}, which vanish in the Euler-Bernoulli beam theory, i.e., $\varepsilon_{22}=0\,,$ $a^{22}b_{22}=0$~\cite{bauchau2009euler}. Finally, Eqs. \ref{eq.8}--\ref{eq.10} simplify the stretching and bending contents to,
    \begin{align}
    E_s = & \frac{Yh}{2} ( {a}^{11} \varepsilon_{11} )^2 + \mathcal{O}(\gamma^4, \delta^4)\,, \label{eq.11} \\
    E_b = & \frac{Y h^3}{24} \left\{ \left[ a^{11}(b_{11}-\bar{b}_{11}) \right]^2\right. \nonumber \\& \left.+ \frac{2}{1+\nu} \left[ a^{11} a^{22} (b_{12}-\bar{b}_{12})^2 \right] \right\}\,, \label{eq.12}
    \end{align}
    leaving only the terms quadratic in strain and moments. 
    
    With Eqs.~\ref{eq.11} and~\ref{eq.12} at hand, we can now derive a reduced 1D theory by coarse-graining the surface metric and second fundamental form tensors about the midcurve at $t=0$. To this end, we calculate the leading order $t$-dependence of the components in $a_{\alpha \beta}, b_{\alpha \beta}, \bar{a}_{\alpha \beta}, \bar{b}_{\alpha \beta}\,,$ followed by an integration over the thin dimension of the 2D energy functional (Eqs. \ref{eq.4},~\ref{eq.11},~\ref{eq.12}) to derive an effective 1D energy. We begin by expanding the position vector of the surface $\mathbf{X}(\sigma, t)$ into Taylor series in powers of $t$,    
    \begin{gather}
    \mathbf{X}(\sigma, t)
    = \mathbf{X}(\sigma,0) + t \frac{\partial \mathbf{X}}{\partial t}\bigg|_{t=0} + \frac{t^2}{2} \frac{\partial^2 \mathbf{X}}{\partial t^2}\bigg|_{t=0} + \mathcal{O}(t^3)\,, \nonumber \\
    \equiv \mathbf{R}(\sigma) + t \; U(\sigma) \mathbf{\hat{n}}(\sigma) +\frac{t^2}{2}\mathbf{P}(\sigma) + \mathcal{O}(t^3)\,, \label{eq.13}
    \end{gather} 
    where, in the current configuration, $\mathbf{R}$ is the position vector of the midcurve. The scalar metric $g$ (Fig.~\ref{fig:coarse}e) defines the unit tangent vector of the curve $\partial_s\mathbf{R}$ through an arc length parametrization of the current configuration $s$
    \begin{gather}
        \partial_1 \mathbf{R}\equiv \frac{\partial \mathbf{R}}{\partial \sigma} = \sqrt{g} \frac{\partial \mathbf{R}}{\partial s}\,. \label{eq.14}
    \end{gather} 
    We also identify the transverse surface curvature vector of the midcurve $\mathbf{P}$ and $\mathbf{\hat{n}}$ (the normal vector to the curve on the surface) by defining the square root of the metric along the $t$ direction as $U$, 
    \begin{gather}
    \frac{\partial \mathbf{X}}{\partial t}\bigg|_{t=0} = U \mathbf{\hat{n}}\,, \\
    \frac{\partial^2 \mathbf{X}}{\partial t^2}\bigg|_{t=0} = \mathbf{P}\,.
    \label{eq.15} 
    \end{gather}
    The normal vector of the surface $\mathbf{\mathbf{\hat{N}}} = \partial_s\mathbf{R} \times \hat{\mathbf{n}}$ completes the triad of the Darboux vectors along the surface curve that satisfy the Darboux equations for the current configuration (cf. Eq.~\ref{eq:Darboux}). 

    The first and second derivatives of Eq.~\ref{eq.13} yield the series expansion of the metric and second fundamental form tensors $a_{\alpha\beta}, b_{\alpha\beta}$ across the strip on the surface. This also gives the explicit $t$-dependence of the tensor components, making them amenable to integration over $t\in\left[-W/2, W/2\right]$ to find a coarse-grained curve that represents the strip in the current configuration (Fig.~\ref{fig:coarse}d). Using Eq.~\ref{eq.8} that gives $a_{12}=\bar{a}_{12}=0 + \mathcal{O}(h^2)$ and Eq.~\ref{eq.9} that leads to $a_{22} = U^2 = 1 + \mathcal{O}(\nu \delta)$ when a  semigeodesic form of the reference metric $\bar{a}_{22}=1$ is used, the metric tensor and the second fundamental form of the surface strip in the current configuration become ($k_g, k_N, \tau_g:$ the geodesic curvature, normal curvature and geodesic torsion of the midcurve, respectively as depicted in Fig.~\ref{fig:coarse}c--e, $k_{N,2}:$ second normal curvature, $K\equiv k_N k_{N,2}-\tau_g^2:$ Gaussian curvature)    
    \begin{gather}
    \mathbf{a} = \begin{bmatrix}
    g\left[\left(1-t\;k_g\right)^2 - t^2 \; K \right] + \mathcal{O}(t^3) & 0\\ 
    0 & 1 + \mathcal{O}(\delta, t)
    \end{bmatrix} ,   \label{eq.16}     
    \end{gather}
    \begin{gather}
    \mathbf{b} = \begin{bmatrix}
    g \; k_N  & \sqrt{g} \; \tau_g\\ 
    \sqrt{g} \; \tau_g & k_{N,2}
    \end{bmatrix}  + \mathcal{O}(t)\,. \label{eq.17}
    \end{gather}
    We define the second normal curvature of the midcurve in the transverse $t$-direction as $k_{N,2}\equiv\frac{1}{U^2}\partial^2_{tt}\mathbf{X}|_{t=0}\cdot\mathbf{\mathbf{\hat{N}}}\,.$ To arrive at Eqs.~\ref{eq.16},~\ref{eq.17}, we have also used the following identities ($U=1$),
    \begin{gather}
    \partial_{s}\mathbf{R} \cdot \partial_s\mathbf{\hat{n}} = -k_{g}, \nonumber\\
    \partial_{s}\mathbf{\hat{n}} \cdot \partial_{s}\mathbf{\hat{n}} = k^2_g + \tau^2_{g}, \\
    \partial_{s}\mathbf{R} \cdot \partial_{s}\mathbf{P} = - k_{N}k_{N,2}. \nonumber 
    \end{gather}
    Similarly, to express the reference tensor coefficients in Eqs.~\ref{eq.11}--\ref{eq.12} in terms of the geometric variables of the reference curve, we next expand the position vector of the reference curve on the surface into Taylor series about a point $t=0$ as,
    \begin{equation} 
    \mathbf{\bar{X}}(\sigma, t) 
    = \mathbf{\bar{R}}(\sigma) + t \; \mathbf{\hat{\bar{n}}}(\sigma) +\frac{t^2}{2}\mathbf{\bar{P}}(\sigma) + \mathcal{O}(t^3), \label{eq.18}
    \end{equation} 
    where, $\mathbf{\bar{R}}, \mathbf{\hat{\bar{n}}}, \mathbf{\bar{P}}$ correspond to the quantities in the reference configuration, which may not be compatible with the surface at all points. Because the energy is invariant under different coordinate representations, we choose to express the variables associated with the reference curve in terms of the current arc length $s$ over the reference arc length $\bar{s}$ for convenience.  Since the arc length $s$ defined for the midcurve in the current configuration (Fig.~\ref{fig:coarse}c), we introduce the relation between $s$ and the arc length $\bar{s}$ of the reference curve
    \begin{equation}
    \partial_s=\sqrt{\frac{\bar{g}}{g}}\partial_{\bar{s}}\,.
    \label{eq:arc_length_transform}    
    \end{equation}
    Then, the tangent vector of the reference curve is transformed through Eq.~\ref{eq:arc_length_transform} as (cf. Eq.~\ref{eq:ref_arc_length})
    \begin{equation}
        \partial_s \mathbf{\bar{R}} = \frac{1}{\sqrt{g} }\partial_\sigma \mathbf{\bar{R}}=\frac{\sqrt{\bar{g}}}{\sqrt{g} }\partial_{\bar{s}} \mathbf{\bar{R}}\,,    \label{eq.20}     
    \end{equation}
    which leads to, 
    \begin{equation}
        \partial_s \mathbf{\bar{R}} \cdot \partial_s \mathbf{\bar{R}}  = \frac{\bar{g}}{g}\quad\text{since}\quad |\partial_{\bar{s}}\mathbf{\bar{R}}|=1\,. 
    \end{equation}
    Defining $\Lambda\equiv\sqrt{g/\bar{g}}$ as the stretch, we use Eqs.~\ref{eq:arc_length_transform} and~\ref{eq.20} to express the curvatures and the geodesic torsion of the reference curve given in Eq.~\ref{eq:Darboux} in the coordinate of the arc length $s$: $\bar{k}_g=\Lambda^2\partial^2_{ss}\mathbf{\bar{R}}\cdot\mathbf{\hat{\bar{n}}}\,,\; \bar{k}_N=\Lambda^2\partial^2_{ss}\mathbf{\bar{R}}\cdot\mathbf{\hat{\bar{N}}}\,,\; \bar{\tau}_g=\Lambda\partial_{s}\mathbf{\hat{\bar{n}}}\cdot\mathbf{\hat{\bar{N}}}$. The surface metric tensor $\mathbf{\bar{a}}$ and the second fundamental form $\mathbf{\bar{b}}$ at a position on the reference curve are then given by ($\bar{K}\equiv \bar{k}_N \bar{k}_{N,2} - \bar{\tau}^2_g:$ Gaussian curvature of the surface defined at a point on the reference curve),
    \begin{gather}
    \mathbf{\bar{a}} = \begin{bmatrix}
    g\left[ \left(\frac{1}{\Lambda^2}-1\right)+  \left(1-t\;\frac{\bar{k}_g}{\Lambda^2}\right)^2 - t^2 \; \bar{K} \right] + \mathcal{O}(t^3) & 0\\ 
    0 & 1
    \end{bmatrix}  ,  \label{eq.21}     
    \end{gather}
    \begin{gather}
    \mathbf{\bar{b}} = \begin{bmatrix}
    \frac{g}{\Lambda^2} \bar{k}_N  & \frac{\sqrt{g}}{\Lambda} \bar{\tau}_g\\ 
    \frac{\sqrt{g}}{\Lambda}\bar{\tau}_g & \bar{k}_{N,2}
    \end{bmatrix}    + \mathcal{O}(t)\,. \label{eq.22}
    \end{gather}
    Combining Eqs.~\ref{eq.4},~\ref{eq.11},~\ref{eq.12},~\ref{eq.16},~\ref{eq.17},~\ref{eq.21},~\ref{eq.22} and integrating Eqs.~\ref{eq.4} over the thin dimension $t \in [-W/2,W/2]\,,$ we arrive at the reduced 1D energy functional for elastic rods
    \begin{gather}
        E_{1D} = \frac{Y}{8} \int \sqrt{\bar{g}} \; d\sigma \left\{ hW \left( 1 - \frac{1}{\Lambda^2} \right)^2 \right. \nonumber  \\
         \left. + \frac{h^3W}{3} \left[ \left(k_N - \frac{1}{\Lambda^2}\bar{k}_N \right)^2 + \frac{2}{1+\nu} \left(\tau_g - \frac{1}{\Lambda}\bar{\tau}_g\right)^2 \right] \right. \nonumber \\ 
         \left. + \frac{hW^3}{3} \left(k_g - \frac{1}{\Lambda^2} \bar{k}_g \right)^2 + \mathcal{O}(\gamma^2 \delta, \gamma^2 \delta^2) \right\}\,. \label{eq.23a}   
    \end{gather}
    We have again neglected the contributions of order three and higher by imposing the small strain approximation. Defining $\ell$ as the length of the reference curve, we nondimensionalize Eq.~\ref{eq.23a} such that $ \sqrt{g}\rightarrow \ell{\sqrt{g}}\,,$ $k_g\rightarrow k_g/\ell\,,$ $k_N\rightarrow k_N/\ell\,,$ $\tau_g\rightarrow \tau_g/\ell\,,$ and likewise for their reference counterparts. By substituting Eq.~\ref{eq.27},~\ref{eq.28},~\ref{eq.29} for $\bar{k}_g\,,$ $\bar{k}_N\,,$ $\bar{\tau}_g\,,$ the unitless form of Eq.~\ref{eq.23a} becomes in terms of the energy units of $\Sigma \equiv YhW\ell/8$ and aspect ratio $\gamma\equiv h/\ell=W/\ell$, 
    \begin{gather}
        w_{1D} \equiv \left( 1 - \frac{1}{\Lambda^2} \right)^2 + \frac{\gamma^2}{3} \left\{ \left[k_g - \frac{\bar{k}\cos\phi}{\Lambda^2}\right]^2\right. \nonumber \\ + \left[k_N - \frac{\bar{k}\sin\phi}{\Lambda^2} \right]^2
        \left.  + \frac{2}{1+\nu} \left[\tau_g - \frac{1}{\Lambda}\left(\bar{\tau}-\Lambda\frac{\partial\phi}{\partial s}\right)\right]^2 \right\}\nonumber\\ + \mathcal{O}(\gamma^2 \delta, \gamma^2 \delta^2)\,, \label{eq.23}  \\  
        \Rightarrow E_{1D}= \int \sqrt{\bar{g}} \; d\sigma \;w_{1D}\,. \nonumber 
    \end{gather}   
    We note that the choice of the current metric as the measure of the elasticity tensor in Eq.~\ref{eq:elasticity_tensor} leads to a stretching term in the Almansi strain measure with a physical behavior that restricts indefinite compression. Another consequence of our constitutive model is that the bending energy has the intuitive pure curvature terms from Eq.~\ref{eq.2} (see Appendix \ref{ap E} for another constitutive model for comparison). The sole role of the $\Lambda$-dependent factors that appear in the bending terms in Eqs.~\ref{eq.23a},~\ref{eq.23} is to rescale the corresponding reference geometric variables in the frame of the current configuration, so, they do not introduce a multiplicative coupling between stretching and bending. Importantly, although $\bar{k}$ and $\bar{\tau}$ are fixed for the reference curve, its transformed geometric variables on the surface $\bar{k}_g$, $\bar{k}_N$, $\bar{\tau}_g$ can change because of the twist degree of freedom manifested by the rotation angle $\phi$ (Fig.~\ref{fig:DtoF}). 
    
    Because of the derivation of the theory from first principles, all elastic moduli are explicitly given by geometric and physical parameters: for a beam with a rectangular cross-section, a stretching modulus can be defined as $A_0=YhW/8$ and the bending moduli become $A_1=YhW^3/24\,,$ $A_2=Y h^3 W/24\,,$ and $A_3=Y h^3 W/(12(1+\nu))$ (cf. Eq.~\ref{eq.2} and Eq.~\ref{eq.23a}). Among these constants, $A_1$ and $A_2$ can immediately be recognized as $A_1=A_2=Y I/2\,,$ where $I\equiv hW^3/12$ is the area moment of inertia for a rectangular cross-section, in agreement with the bending of an Euler-Bernoulli beam. The relation of $A_3$ to the Poisson's ratio $\nu$ is less straightforward. Altogether, the primary contribution of this work is Eq.~\ref{eq.23a}, which provides an elastic energy functional for extensible slender objects modeled as 1D elastic curves. This will allow us to formulate the problem of an elastic curve (with a reference curvature and torsion) bounded to a rigid surface where it will be subject to additional compatibility constraints.

\section{Elastic Curves Bounded on Surfaces}
\label{sec.4}
    \subsection{Surface constraints}
    When an elastic curve is confined on a rigid background surface, the second fundamental form of the current configuration (Eq.~\ref{eq.17})  will be constrained by the surface invariants since every point on the curve must acquire the normal vector to the surface $\mathbf{\hat{N}}$ at that point. These two invariants are the mean curvature $H$ and the Gaussian curvature $K$, which can be given as functions of the surface parametrization $x^1\equiv u\,, x^2\equiv v\,.$ So, the two constraint equations along the  curve parametrized by $\sigma$ are
    \begin{gather}
        H(u(\sigma),v(\sigma)) = \frac{1}{2}\left[k_N(\sigma) + k_{N,2}(\sigma)\right]\,, \label{eq.24} \\
        K\left(u(\sigma),v(\sigma)\right) = k_N(\sigma)k_{N,2}(\sigma) - \tau_g^2(\sigma)\,. \label{eq.25}
    \end{gather}
    Using Eqs.~\ref{eq.23a},~\ref{eq.24},~\ref{eq.25}, we define the constrained unitless energy functional for a curve on a rigid surface ($\lambda_1, \lambda_2:$ Lagrange multipliers),
    \begin{gather}
        w^C_{1D} \equiv  \; \left\{ w_{1D}  + \lambda_1\left[K-\left(k_Nk_{N,2}-\tau_g^2\right) \right] \right.\nonumber \\
        \left. +\lambda_2 \left[2H-(k_N+k_{N.2})\right] \right\}\,,  \nonumber \\
        \text{and}\quad E^C_{1D} \equiv \int \sqrt{\bar{g}} \, d\sigma \, w^C_{1D}\,. \label{eq.26}
    \end{gather}
    Eq.~\ref{eq.26} can be written in a closed form as a functional of eight curve variables $\mathbf{f}\equiv\{g, k_g, k_N, k_{N,2}, \tau_g, \phi, \lambda_1, \lambda_2\}\,.$
    In addition to the variation of Eq.~\ref{eq.26} that gives eight Euler-Lagrange (EL) equations for the eight unknowns of the current curve configuration at equilibrium, there is also dependence of the surface coordinates $u, v$ on $\sigma$ in the constraints given by Eq.~\ref{eq.24} and~\ref{eq.25} for surfaces with nonconstant $H$ or $K$, or both. That is, as the curve is varied on a nonuniform surface, the constraints coming from the background surface change with $u, v$. So, we would need two more equations to simultaneously solve for $u(\sigma), v(\sigma)$ along the curve to determine the curve position in $\mathcal{S}\,.$ In what follows, we will first derive the EL equations and then the mapping equations to fully determine the problem.

    \subsection{Euler-Lagrange equations for the equilibrium curves}

    When bounded on a curved surface, a reference curve specified by $\bar{g}\,, \bar{k}\,, \bar{\tau}$ would minimize the energy in Eq.~\ref{eq.26} by stretching, bending and twisting. To find the current equilibrium configuration of the curve on the surface, we derive the Euler-Lagrange equations and the corresponding natural boundary conditions. By defining $\omega_{,f} \equiv \partial \omega^C_{1D}/\partial f_n$ for any general curve variable $f_n$ ($n=1\dots8$), the variation of Eq.~\ref{eq.26} is
    \begin{gather}
    \delta E^C_{1D} =  \int \;  \; \sqrt{\bar{g}} \; d\sigma \left( \omega_{,g} \; \delta g + \omega_{,k_g} \; \delta k_{g} \right. \nonumber \\
    +\;\omega_{,k_N} \; \delta k_{N} +\;\omega_{,k_{N,2}} \; \delta k_{N,2} + \omega_{,\tau_g} \; \delta \tau_{g}  \nonumber \\
    \left.+ \omega_{,\phi} \; \delta \phi + \omega_{,\lambda_1} \; \delta \lambda_1 + \omega_{,\lambda_2} \; \delta \lambda_2  \right). \label{eq.31}
    \end{gather}
    We convert the variation $\delta f_n$ into the independent variations of the position vector of the curve $\delta\mathbf{X(\sigma)}$ and the rotations of the Darboux frame $\delta\phi(\sigma)$, so that they give a total of 8 Euler-Lagrange equations for the 8 unknowns. This approach removes the need to impose the surface compatibility conditions known as the Gauss-Codazzi-Mainardi equations as additional constraints to Eq.~\ref{eq.26}~\cite{efrati2009elastic}.
    
    To write the Euler-Lagrange equations in a compact form, we define the equilibrium tensions $F_g\,, F_{k_g}\,, F_{k_N}$ along the three Darboux vectors $\partial_s\mathbf{R}, \mathbf{\hat{n}}, \mathbf{\hat{N}}\,,$ respectively, as
    \begin{gather}
    F_{g} \equiv   \sqrt{\frac{1}{g}} \left[ 2 g w_{, g} - (k_g w_{, k_g} + k_N w_{, k_N}  + \tau_g w_{, \tau_g} )  \right]\,, \label{eq.39} \\ 
    F_{k_g} \equiv \partial_\sigma \left( \frac{w_{,k_g}}{g} \right) - \sqrt{\frac{1}{g}} \left( w_{,k_N} \tau_g - w_{,\tau_g} k_N\right)\,, \label{eq.40} \\
    F_{k_N} \equiv \partial_\sigma \left( \frac{w_{,k_N}}{g} \right) + \sqrt{\frac{1}{g}} \left( w_{,k_g} \tau_g - w_{,\tau_g} k_g\right)\,. \label{eq.41}
    \end{gather}
    Then, alongside the two constraints in Eqs.~\ref{eq.24},~\ref{eq.25}, the Euler-Lagrange equations of Eq.~\ref{eq.31} are ($w_{, \bar{f}}\equiv \partial w_{1D}/\partial \bar{f}=\partial_{\phi} w_{1D}\left(\partial_{\phi} \bar{f}\right)^{-1},$ where $\bar{f}\equiv \{\bar{k}_g\,, \bar{k}_N, \bar{\tau}_g\}$),
    \begin{gather}
        \partial_\sigma F_{g} + \sqrt{g} \; k_g \; F_{k_g} + \sqrt{g}\;  k_N \; F_{k_N} = 0,  \label{eq.32} \\
        \partial_\sigma F_{k_g} - \sqrt{g} \;   k_g F_{g} - \sqrt{g} \; \tau_g \; F_{k_N} = 0,     \label{eq.33} \\
        \partial_\sigma F_{k_N} - \sqrt{g} \;  k_N F_{g} + \sqrt{g} \; \tau_g \; F_{k_g} = 0,  \label{eq.34} \\
        \partial_\sigma \left(\sqrt{\frac{1}{g}} w_{,\bar{\tau}_g} \right) + \left( \bar{k}_g w_{,\bar{k}_N} - \bar{k}_N w_{,\bar{k}_g}\right) = 0,   \label{eq.35} \\        
        \lambda_2 + k_N \; \lambda_1 = 0, \label{eq.36}  \\    
        \partial_\sigma \left(\sqrt{\frac{1}{g}} \omega_{,\tau_g} \right) + \left( k_g \;\omega_{,k_N} - k_N \;\omega_{,k_g}\right) = 0.        \label{eq.37}        
    \end{gather}
    
    We identify Eqs.~\ref{eq.32}--\ref{eq.34} as the force balance equations for a stretchable, elastic  rod \cite{starostin2007shape},  which can be compactly written as rotations in $SO(3)$ as 
    \begin{gather}
        \partial_s \mathbf{F} = \mathbf{\Omega} \times \mathbf{F}, 
    \end{gather}
    such that,
    \begin{gather}
        \mathbf{\Omega} \equiv \left[ -\tau_g, -k_N, k_g \right]\,, \quad   
        \mathbf{F} \equiv \left[ F_g, F_{k_g},F_{k_N} \right]\,.  
    \end{gather} 
    Furthermore, it is evident that the torsional moment balance equation in Eq.~\ref{eq.35} arises from the variation of $\phi$. 
    
    Simplifying Eq.~\ref{eq.37} using Eq.~\ref{eq.35} returns a partial differential equation (PDE) first order in $\lambda_1$,
    \begin{equation}
        \partial_\sigma\left(\frac{2}{\sqrt{g}} \lambda_1 \tau_g\right) - k_g (\lambda_1 k_{N,2} + \lambda_2) =0\,. \label{eq.42}
    \end{equation}
    Eqs.~\ref{eq.36} and~\ref{eq.42} together solve for the Lagrange multipliers $\lambda_1, \lambda_2$. Furthermore, substituting Eq.~\ref{eq.37} into Eq.~\ref{eq.32} yields a simplified PDE first order in $g$,
    \begin{equation}
    \begin{aligned}
    &\partial_\sigma\left\{ \sqrt{\frac{1}{g}} \left[ 2 g w_{,g} - \left(k_g w_{,k_g} + k_N w_{,k_N}  \right) \right]\right\} - \sqrt{\frac{1}{g}}(\partial_\sigma \tau_g) w_{\tau_g} \\
    & + \tau_g\left( k_g \;\omega_{,k_N} - k_N \;\omega_{,k_g}\right) + \sqrt{g} k_g \; F_{k_g} + \sqrt{g} k_N \; F_{k_N} = 0\,. 
    \end{aligned}
    \label{eq.43}       
    \end{equation}
    Since Eq.~\ref{eq.43} is first order in $\sigma\,,$ it requires one clamped boundary condition $g=g_0$ to obtain a solution. 
    Additionally, the following natural (Neumann) boundary conditions hold for Eqs.~\ref{eq.33}--\ref{eq.35}, \ref{eq.42}, respectively,
    \begin{gather}
        F_{k_g}|_{\sigma=0} = F_{k_g}|_{\sigma=1} = 0\,, \label{eq.38a} \\        F_{k_N}|_{\sigma=0} = F_{k_N}|_{\sigma=1} = 0\,,  \\
        w_{,\bar{\tau}_g}|_{\sigma=0} =  w_{,\bar{\tau}_g}|_{\sigma=1} = 0\,,  \\
        w_{,\tau_g}|_{\sigma=0} =  w_{,\tau_g}|_{\sigma=1}\,. \label{eq.38b}
    \end{gather}
    \subsection{Mapping equations between the curve and the surface}
    
    To complete the formulation, we next derive two equations for $u(\sigma)$ and $v(\sigma)$, which depend on the local features of the surface. These two equations will be given in unitless form by introducing a ratio $\zeta \equiv \ell/D$ between the length scale of the curve $\ell$ ($\sqrt{g}\rightarrow \ell \sqrt{g}$) and that associated with the surface $D$ ($a_{\alpha\beta} \rightarrow D^2 a_{\alpha\beta}$). The parameter $\zeta$ determines the normalized length of the curve on the surface. For a general curve (parametrized by $\sigma$) on a surface (parametrized by the two coordinates $u, v$), the following equations hold as derived in Appendix~\ref{ap F}:
    \begin{gather}
        g(\sigma) = \frac{1}{\zeta^2} \left[ (\partial_\sigma u)^2 \; a_{uu} + 2\;(\partial_\sigma u) (\partial_\sigma v) \; a_{uv} + (\partial_\sigma v)^2 \; a_{vv} \right]\,, \label{eq.44} \\
        k_g(\sigma) = \frac{1}{(g\;\zeta^2)^{3/2}}\sqrt{|a|} \bigg[(\partial_\sigma u)^3\;\Gamma^v_{uu} - (\partial_\sigma v)^3\;\Gamma^u_{vv} \nonumber \\ 
        + (\partial_\sigma u)^2(\partial_\sigma v)\;(2\Gamma^v_{uv}-\Gamma^u_{uu}) 
        - (\partial_\sigma v)^2(\partial_\sigma u)\;(2\Gamma^u_{uv}-\Gamma^v_{vv}) \nonumber \\  
        + (\partial_{\sigma\sigma} u)(\partial_\sigma v) - (\partial_{\sigma\sigma} v)(\partial_\sigma u) \bigg]\,,  \label{eq.45} \\
        k_N(\sigma) = \frac{1}{g\;\zeta^2}\left[(\partial_\sigma u)^2 \; b_{uu} + 2\;(\partial_\sigma u) (\partial_\sigma v) \; b_{uv} + (\partial_\sigma v)^2 \; b_{vv} \right]\,,  \label{eq.46} \\
        k_{N,2}(\sigma) = \frac{1}{|a|\;g\;\zeta^2} \bigg[ \big( a_{uu}\;(\partial_\sigma u) + a_{uv}\;(\partial_\sigma v) \big)^2 b_{vv} \nonumber \\ - 2\;\big(a_{uu}\;(\partial_\sigma u) + a_{uv}\;(\partial_\sigma v)\big)\big(a_{uv}\;(\partial_\sigma u) + a_{vv}\;(\partial_\sigma v)\big) \; b_{uv} \nonumber \\
        + \big( a_{uv}\;(\partial_\sigma u) + a_{vv}\;(\partial_\sigma v) \big)^2 b_{uu} \bigg]\,,  \label{eq.47} \\
        \tau_g(\sigma) = \frac{1}{\sqrt{|a|}\;g\;\zeta^2}\bigg[ (\partial_\sigma u)^2\left( b_{uv}a_{uu}-b_{uu}a_{uv} \right) \nonumber \\ 
        + (\partial_\sigma v)^2\left( b_{vv}a_{uv}-b_{uv}a_{vv} \right) \nonumber \\
        + (\partial_\sigma u)(\partial_\sigma v)\left( b_{vv}a_{uu}-b_{uu}a_{vv} \right) \bigg]\,.  \label{eq.48}
    \end{gather}
    In Eqs.~\ref{eq.44}--\ref{eq.48}, the elements of the metric tensor $a_{\alpha\beta}\,,$ the second fundamental form $b_{\alpha\beta}\,,$ and the Christoffel symbols $\Gamma^{\gamma}_{\alpha\beta}$  can be calculated from Eq.~\ref{eq:2nd_deriv}. Then, along with Eqs.~\ref{eq.24},~\ref{eq.25},~\ref{eq.39}--\ref{eq.41},~\ref{eq.33}--\ref{eq.36},~\ref{eq.42} and~\ref{eq.43}, we can solve for $u(\sigma), v(\sigma)$ from a subset of the five geometric relations given by Eqs.~\ref{eq.44}--\ref{eq.48} when two boundary conditions $u_0$ and $v_0$ for the position of the clamped end of the curve are specified on the surface. 

    To determine two independent equations out of Eqs.~\ref{eq.44}--\ref{eq.48}, we first recall the surface constraints on the curve: when the mean and Gaussian curvatures $H(u,v), K(u,v)$ are given for a rigid surface, it can easily be shown that only one of Eqs.~\ref{eq.46}--\ref{eq.48} is independent per Eqs.~\ref{eq.24},~\ref{eq.25}.
    Similarly, for a given surface, only one of Eqs.~\ref{eq.44} and~\ref{eq.45} is independent. To see this, we write Eq.~\ref{eq.44} in a compact form in the arc length coordinate as, 
    \begin{equation}
        1 = a_{\alpha\beta} \; \partial_s w^{\alpha} \partial_s w^{\beta}\,, \label{eq.56a}
    \end{equation}
    where $\textbf{w}=(u,v)$. Taking the derivative of Eq.~\ref{eq.56a} gives,
    \begin{gather}
        0 = a_{\alpha\beta} \; \partial_s w^{\alpha} (\partial_{ss} w^{\alpha} + \Gamma^\alpha_{\gamma \sigma} \partial_s w^{\gamma} \partial_s w^{\sigma}), \nonumber \\
        = a_{\alpha\beta} \; \partial_s w^{\alpha} D_s (\partial_s w^{\beta}), 
        \nonumber \\
        = \partial_s \mathbf{w} \cdot D_s (\partial_s \mathbf{w})\,,
        \label{eq.57a}
    \end{gather}
    where we have introduced the covariant derivative along the curve $D_s$ on the surface, so that $D_s (\partial_s \mathbf{w}$) is the covariant acceleration whose component along the curve tangent vanishes per Eq.~\ref{eq.57a}. Then, we can define $k_g$ as the deviation of the curve, normal to the curve tangent vector $\partial_s \mathbf{w^{\perp}} \equiv (\partial_s v, - \partial_s u)$, (i.e., in $\mathbf{\hat{n}}$ direction), so that,

    \begin{gather}
    k_g \equiv \frac{1}{\sqrt{|a|}} \left( \partial_s\mathbf{w^{\perp}} \cdot D_s (\partial_s\mathbf{w}) \right),          
    \end{gather}
    which leads to Eq.~\ref{eq.45}. Therefore, we only need to use one equation each from Eqs.~\ref{eq.44}--\ref{eq.45} and Eqs.~\ref{eq.46}--\ref{eq.48}.
    

    With these at hand, we have all the ingredients to solve the following problem: \textit{for a given reference space curve with intrinsic length $\bar{g}$,  curvature $\bar{k}$, and  torsion $\bar{\tau}$, restricted to be on a fixed surface given by $a_{\alpha\beta}, b_{\alpha\beta}$, and tethered to a point $(u_0, v_0)$ on the surface, we will find the current configuration of the surface-bound curve at equilibrium by solving the eight Euler-Lagrange equations (Eqs.~\ref{eq.24},~\ref{eq.25},~\ref{eq.33}--\ref{eq.36},~\ref{eq.42} and~\ref{eq.43}) along with the two mapping equations (one from each subset of Eqs.~\ref{eq.44}--~\ref{eq.45} and ~\ref{eq.46}--~\ref{eq.48}) for the eight unknowns along the curve $(g, k_g, k_N, k_{N,2}, \tau_g, \phi, \lambda_1, \lambda_2)$ and their corresponding mapping $(u(\sigma),v(\sigma))$ on the surface.}

\section{Results}
\label{sec.5}
    \subsection{Simulation procedure}
    
    We simulate the equilibrium energy configurations of inextensible and extensible curves, with an array of curvatures $\bar{\kappa}$ and torsions $\bar{\tau}$ in the reference configuration, on different zero, positive, and negative Gaussian curvature surfaces. For a surface with the parametric equation given by the position vector $\mathbf{X}(u,v)$, the metric $a_{\alpha\beta}$, second fundamental form $b_{\alpha\beta}$, the Christoffel symbols $\Gamma^{\gamma}_{\alpha\beta}$ and in turn the mean and Gaussian curvatures $H, K$ as a function of $(u,v)$ are computed in Wolfram Mathematica 14~\cite{WolframMathematica14}. Then, the weak formulations of Eqs.~\ref{eq.24},~\ref{eq.25},~\ref{eq.33}--\ref{eq.36},~\ref{eq.42}, ~\ref{eq.43},~\ref{eq.44},~\ref{eq.46}, which consist of three algebraic equations and seven nonlinear PDEs,   -- are solved simultaneously by using the FEniCS finite element package on python~3.12 \cite{logg2012automated, alnaes2015fenics, kaplan_1D_elasticity}. Since the set of equations is 10th order in $\sigma\,,$ in addition to the seven natural boundary conditions given in Eqs.~\ref{eq.38a}--\ref{eq.38b}, we impose three Dirichlet boundary conditions on $\sqrt{g}, u, v$ at the tethered end of the curve $\sigma=0$. For nonuniform surfaces, we optimize the tethered point by the downhill simplex method by sweeping different boundary conditions for $u_0, v_0$ to find the equilibrium configuration that locally minimizes the energy on the surface~\cite{nelder1965simplex, wright1996direct}. 
    
    In all computations, the intrinsic curve metric is taken to be $\bar{g} = 1$ everywhere on the curve, and the aspect ratio is set to $\gamma\equiv h/\ell=W/\ell=10^{-1}$. The Poisson's ratio $\nu$ is effectively $0$ for the inextensible cases. For extensible curves, we take $\nu=0.5$ to model an incompressible filament. To visualize the current curve configuration at equilibrium, the prescribed surface and the position of the curve on the surface $\mathbf{X}(u(\sigma),v(\sigma))$ are plotted in Mathematica.

    \subsection{Inextensible curves}

    In this section, we first validate our model through the simulations of inextensible curves on surfaces with constant Gaussian curvature $K$, specifically on a sphere ($K>0$), a cylinder ($K=0$), and a catenoid ($K<0$). Then, we move on to a nontrivial scenario where a general curve is placed on a helicoid, which is a minimal saddle surface ($H=0\,, K<0$).

    
    In the inextensible limit, since $\Lambda = 1$ (i.e., $g=\bar{g}=1$) everywhere on the curve, the stretching term in Eqs.~\ref{eq.25},~\ref{eq.26} vanishes, and Eq.~\ref{eq.2} is recovered. 
    Because $g=\bar{g}=1\,,$ $F_{g}=0$ in Eq.~\ref{eq.39} under the hard inextensibility constraint. In this limit, Eq.~\ref{eq.32} merely 
    reinforces the Darboux frame rotation of $F_{k_g}$ and $F_{k_N}$ in Eqs.~\ref{eq.33},~\ref{eq.34}. Therefore, we only need to solve the other seven Euler-Lagrange equations  (Eqs.~\ref{eq.24},~\ref{eq.25},~\ref{eq.33}--\ref{eq.36},~\ref{eq.42}) and two geometric mapping equations from Eqs. \ref{eq.44}--~\ref{eq.48}.  
    
\subsubsection{Sphere}

      \begin{figure}[t]
        \centering
        \includegraphics[width=\columnwidth] {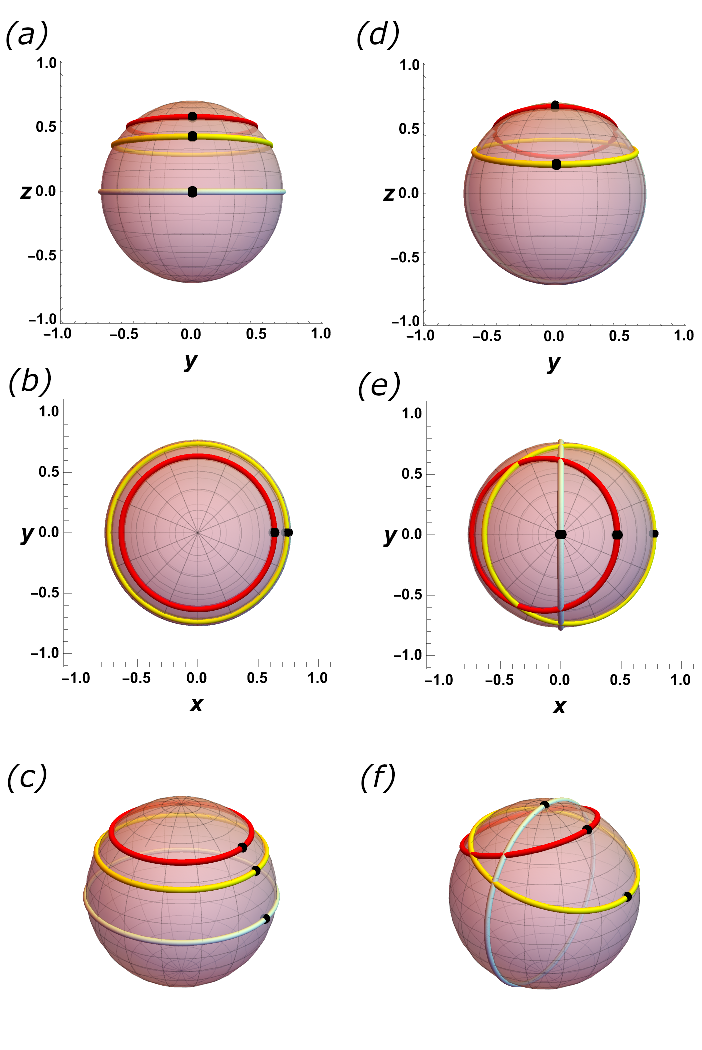}
        \caption{\textbf{Inextensible curves on a unit sphere.} \textbf{(a)} Side, \textbf{(b)} plan,\textbf{ (c)} and elevated views of circles ($\bar{\tau}=0$) with $\bar{k}=1$ (white), and $\bar{k}>1$ (yellow, red). \textbf{(d)--(f)} Since the sphere is an umbilical surface, tethering the curves to another point (black dots) leads to tilted configurations with zero energy. All curves are of normalized length $\zeta = 2\pi$.}
        \label{fig:sphere}
    \end{figure}

    Confinement of filaments, in particular semiflexible polymers, on spherical surfaces has previously been studied theoretically~\cite{ spakowitz2003semiflexible, guven2012confinement}. By using our incompatible theory of 1D elasticity, we here generate results for plane curves ($\bar{\tau}=0$) with a constant reference curvature $\bar{\kappa}$ on a sphere, which has constant positive Gaussian curvature.
    
    The surface of a sphere obeys the parametric equation
    \begin{align}
        \mathbf{X}(u,v) = \big( R \sin{u}\cos{v}, R \sin{u}\sin{v},R \cos{u} \big), \label{eq.51}
    \end{align}
    where $u \in [0, \pi]$ is the polar angle and $v \in [0, 2\pi]$ is the azimuthal angle, and $R$ is the constant radius. Because of symmetry, the relations $k_N=k_{N,2}=\frac{1}{R}$ and $\tau_g=0$ always hold for any curve on the sphere. So, only $k_g$ and $\phi$ can vary along the curve on a spherical surface. Here, we limit our analysis to a unit sphere, i.e., $R=1$.

    Any curve must satisfy the relation 
    \begin{equation}
    \sqrt{k^2_g+k_N^2} = \bar{k}
    \label{eq:totalcurvature}
    \end{equation}
    to remain compatible with the underlying surface. Therefore, planar curves with $\bar{k}<1$ are not compatible with a unit sphere ($R=1, k_N=1$). The deformed configuration of these curves with the lowest energy will assume a geodesic form ($k_g=0$) for $\phi=\pi/2\,,$ running along the equator or any of the meridians (Fig.~\ref{fig:sphere}, white curves). This implies that the only compatible geodesic configuration admissible on a unit sphere is when $\bar{k} = 1$.
    
    When $\bar{k} > 1$, then the curve become compatible with the unit sphere, and the orientation of the curve depends on its placement. That is, with higher $\bar{k}$, the curve will move closer to a chosen pole on the sphere (as an umbilical surface) with increasing $k_g$ (Fig.~\ref{fig:sphere}, red and yellow curves). For example, for a curve with $\bar{\tau}=0\,,\, \bar{k}= 1/\sin{(\pi/3)}>1$, if the tethered point is at $u_0=\frac{\pi}{3}$, then the curve will be a circle of latitude, i.e., $u=\frac{\pi}{3}$ (yellow curve in Fig.~\ref{fig:sphere}a--c). Likewise, if the tethered point of the same curve is moved away from the pole shown in Fig.~\ref{fig:sphere}, it will tilt to match its reference curvature (yellow curve in Fig.~\ref{fig:sphere}d--f). The red curve in Fig.~\ref{fig:sphere} with $\bar{\tau}=0\,,\, \bar{k}= 1/\sin{(\pi/4)}>1$ is another depiction of this case; it becomes a circle of latitude when tethered at $u_0=\frac{\pi}{4}$ and rotates if its tethered point is moved closer to the pole in order to match its constant curvature to the reference curvature (Fig.~\ref{fig:sphere}d--f). 

    \subsubsection{Cylinder}

    \begin{figure*}[ht]
        \centering
        \includegraphics[width=\textwidth]{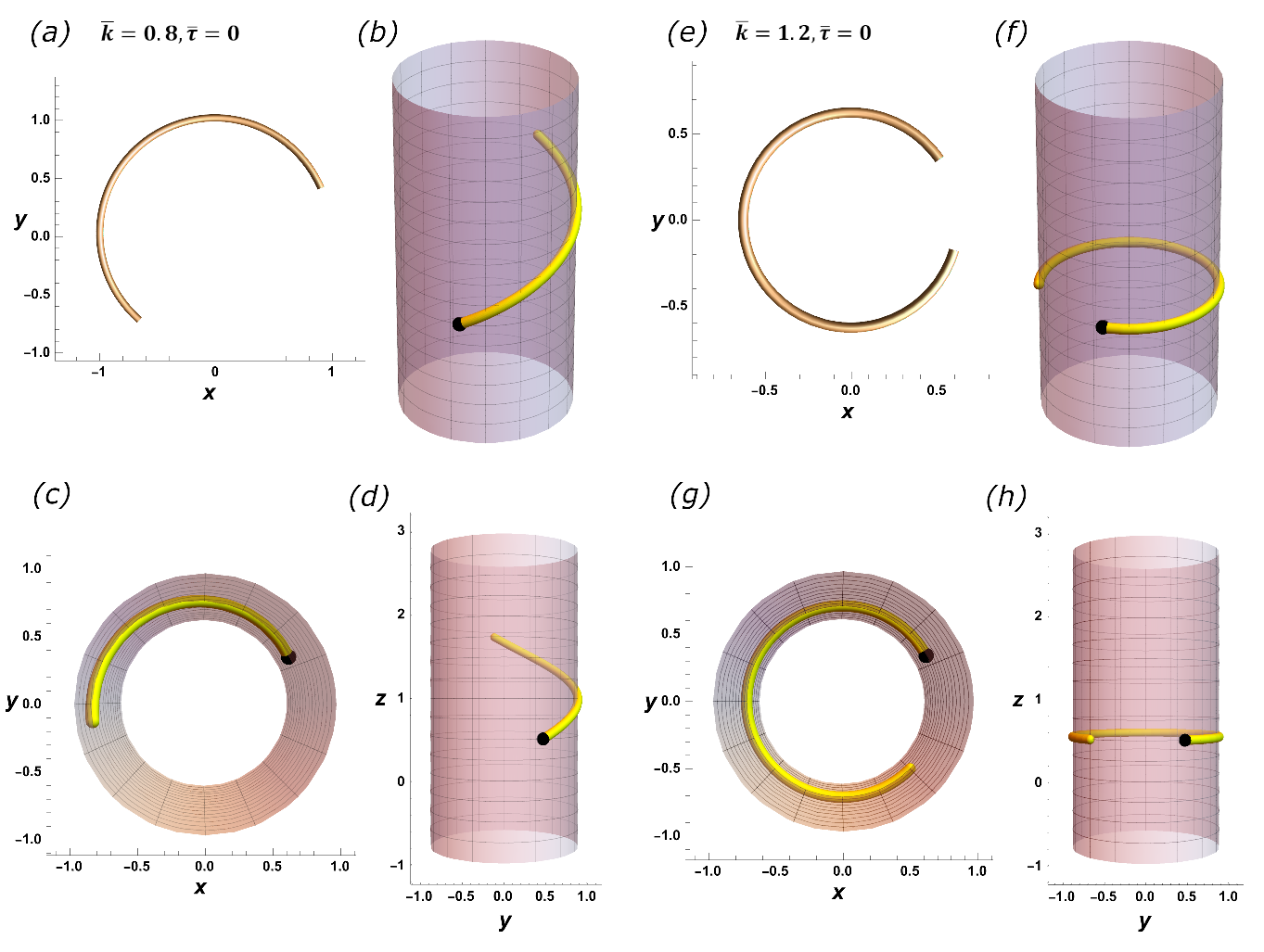}
        \caption{\textbf{Circular arcs placed on a cylinder.}  (a) A planar circular arc ($\bar{\tau}=0$) of normalized length $\zeta=\pi$ with $\bar{k}=0.8\,,$ and \textbf{(b)--(d)} its incompatible configuration with finite $\tau_g\,, \kappa_g\,, \kappa_N$ on the cylinder.  \textbf{(e)} An incompatible planar circular arc of normalized length $\zeta=\pi$ with $\bar{k}=1.2\,,$ and \textbf{(f)--(h)} its finite-energy state on the cylinder. The panels show the  elevated (b, f), plan (c, g), and side views (d, h). The black dots are the tethered points of the curves.}
        \label{fig:cylinder-wrap}
    \end{figure*}  
    
    A cylinder with a unit radius of curvature has the parametrization
    \begin{align}
        \mathbf{X}(u,v) = \big( \cos{v}, \sin{v}, u \big)\,, \label{eq.57}
    \end{align}
    where $u$ is the vertical coordinate, and $v \in[0,2\pi]$ runs along the circumference. From Eq.~\ref{eq.57}, the principal curvature equals $1$ in the $v$ direction and $0$ in the $u$ direction, leading to zero Gaussian curvature. The circular wrapping around the surface ($k_N=1$) and the straight line configuration ($k_N=0$) represent the two planar geodesics ($k_g=0, \tau_g=0$) on the surface. When $\bar{k}\neq1$ for a planar curve, Eq.~\ref{eq:totalcurvature} cannot be satisfied, and the curve can no longer be embedded in the unit cylindrical surface without deformation. This leads to residual stress at equilibrium.

    The equilibrium configurations of the incompatible planar curves on a cylinder with $\bar{k}=0.8<1$ and $\bar{k}=1.2>1$ are shown in Fig.~\ref{fig:cylinder-wrap}a--d and Fig.~\ref{fig:cylinder-wrap}e--h, respectively. To lower its curvature, the curve with $\bar{k}=0.8$ rotates away from the circular wrapping configuration toward the vertical direction and acquires finite $k_g, \tau_g,$ and $k_N<1\,.$ For the curve with $\bar{k}=1.2$, the lowest energy configuration is the geodesic, so that the curve opens up to wrap around the surface in a circular loop with lower curvature (Fig.~\ref{fig:cylinder-wrap}e--h). Thus, the stretching energy is zero, but the curve has bending stress.

    \subsubsection{Catenoid}

        \begin{figure}[t]
        \includegraphics[width= \columnwidth] {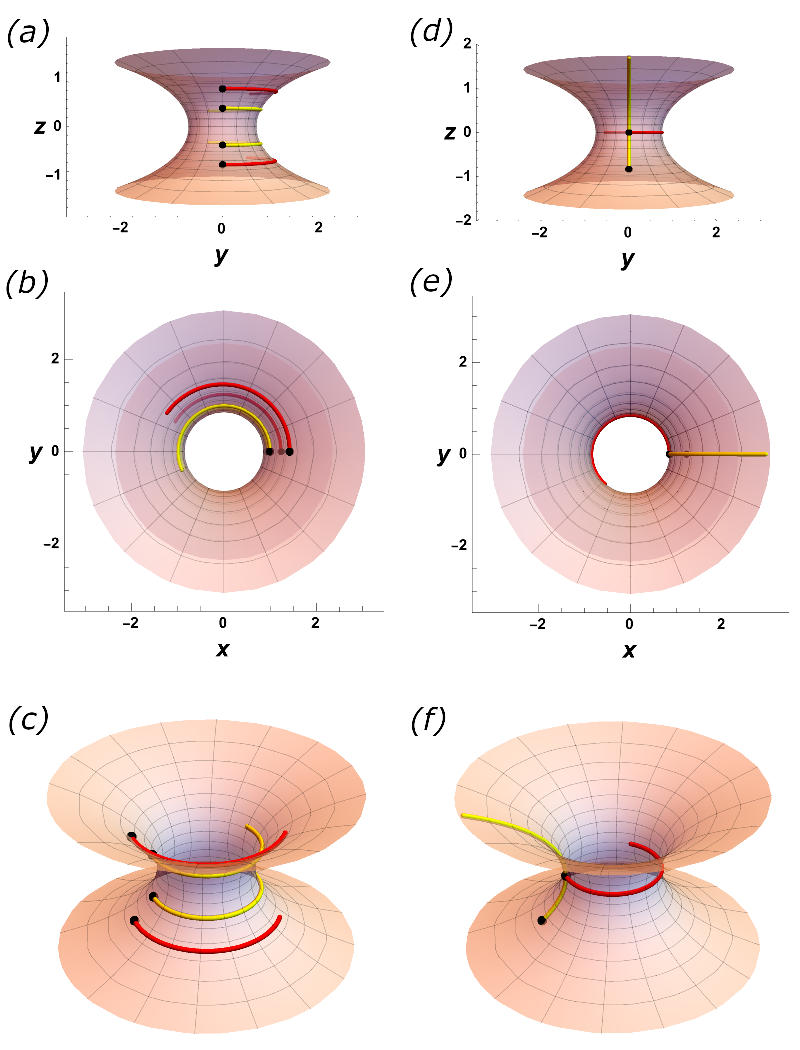}
        \caption{\textbf{Inextensible curves on a catenoid.}
        \textbf{(a)--(c)} Equilibria of planar curves ($\bar{\tau}=0$) with $\bar{k} < 1$ when tethered at one end. The curves above and below the neck of the catenoid have the geodesic curvatures $k_g>0$ and $k_g<0\,,$ respectively. On a catenoid the normal curvature $k_N$ of a curve becomes maximum at the neck. 
        \textbf{(d)--(f)} Incompatible planar curves ($\bar{\tau}=0$) with $\bar{k} \geq 1$ : oriented along the two geodesic directions at equilibrium depending on the position of their tethered end. The residual stresses emerge from the finite normal curvature $k_N\neq \bar{k}\,.$ The figure panels from top to bottom show the side, plan, and elevated views. The black dots are the tethered points of the curves. All curves are of normalized length $\zeta = 4$.}
        \label{fig:catenoid}
    \end{figure}   
    
    Catenoid is a minimal surface ($H=0$) with a negative Gaussian curvature often arising in soft matter systems, such as in capillary bridges, necks connecting two merging membranes in membrane fission during cytokinesis~\cite{powers2002fluid}. The canonical surface parameterization of the catenoid is given by   
    \begin{align}
        \mathbf{X}(u,v) = \left(c \sin{u}\cosh{\left(\frac{v}{c}\right)},\, c\cos{u}\cosh{\left(\frac{v}{c}\right)},\, v \right)\,, \label{eq.64}
    \end{align}  
    where $u \in [0, 2\pi]$ is the periodic parametrization along the horizontal direction, the parametrization along vertical direction $v$ is unbounded in the $\pm z$ direction, and $c$ is the radius of the neck, which we set to $c=1$.
    From the higher order derivatives of Eq.~\ref{eq.64}, the Gaussian curvature of the catenoid is found as $K=-1/\cosh^4{v}$; giving $-1$ on the neck at $v=0$, while its magnitude keeps decreasing in the $\pm z$ direction.

    On a catenoid, planar curves with $\bar{k} < 1$ become compatible away from the neck. In Fig.~\ref{fig:catenoid}a--c, the two yellow curves at $v_0=\pm 0.5$ have the same total curvature $\bar{k}=1/\cosh{(0.5)}$ and differ only by the sign of the geodesic curvature $k_g$ (negative above the neck). If $\bar{k}$ is even smaller, the curves will be located farther away from neck, as depicted by the red curves in Fig.~\ref{fig:catenoid}a--c. 
    
    Setting $\bar{k} \geq 1$ for the reference curve, on the other hand, leads to incompatible planar curve configurations on the catenoid. For $\bar{k} > 1$, the lowest energy deformed configuration is a geodesic around the neck (Fig.~\ref{fig:catenoid}d--f, red curves). Although a reference curve with $\bar{k} = 1$ tethered at $v_0=0$ is compatible and thus yields a ground state (i.e., with zero energy) around the neck (where $k_g = \tau_g = 0,  k_N = 1, u=\sigma, v=0$), it is aligned along the vertical direction in a frustrated (incompatible) state when tethered far away from $v_0=0$, where $k_g = \tau_g= 0, v=\sigma , u=$ constant, and $k_N= -1/\cosh^2{\sigma}$ (Fig.~\ref{fig:catenoid}d--f, yellow curves). In this configuration, the curve is geodesic, but the normal curvature mismatch increases the frustration along the curve with bigger $|v|$.

    \subsubsection{Helicoid}

        \begin{figure}[t]
        \centering
        \includegraphics[width=\columnwidth] {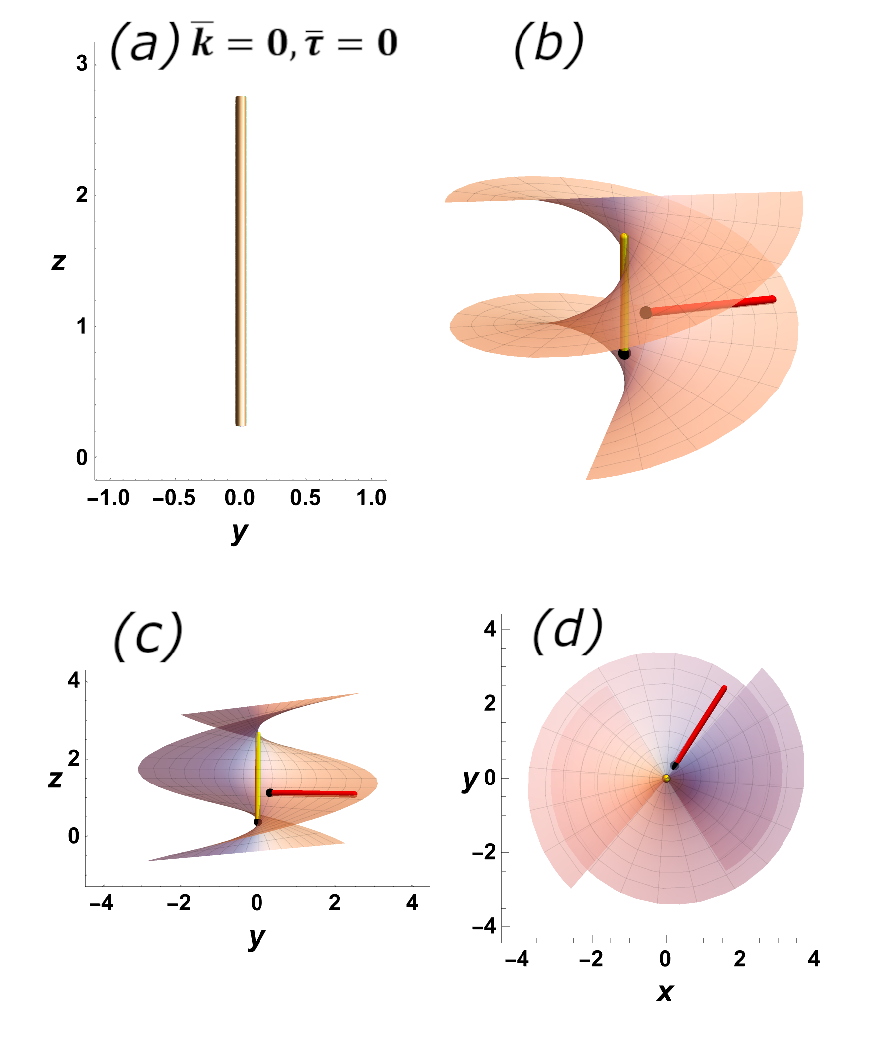}
        \caption{\textbf{Inextensible compatible straight lines on a helicoid.} \textbf{(a)} A straight line ($\bar{\tau}=0$, $\bar{k}=0$) of normalized length $\zeta=3$ as the reference configuration. \textbf{(b)--(d)} Elevated, side, and plan views of the yellow and red lines tethered to a point on the central axis and away from the central axis (black dots), respectively. Both lines align along the two geodesic directions where the finite geodesic torsion $\tau_g$ balances the twist term $\partial\phi/\partial s$ in Eq.~\ref{eq.23}.}
        \label{fig:helicoid-st}
    \end{figure}   
    \begin{figure*}[ht]
        \centering
        \includegraphics[width=\textwidth] {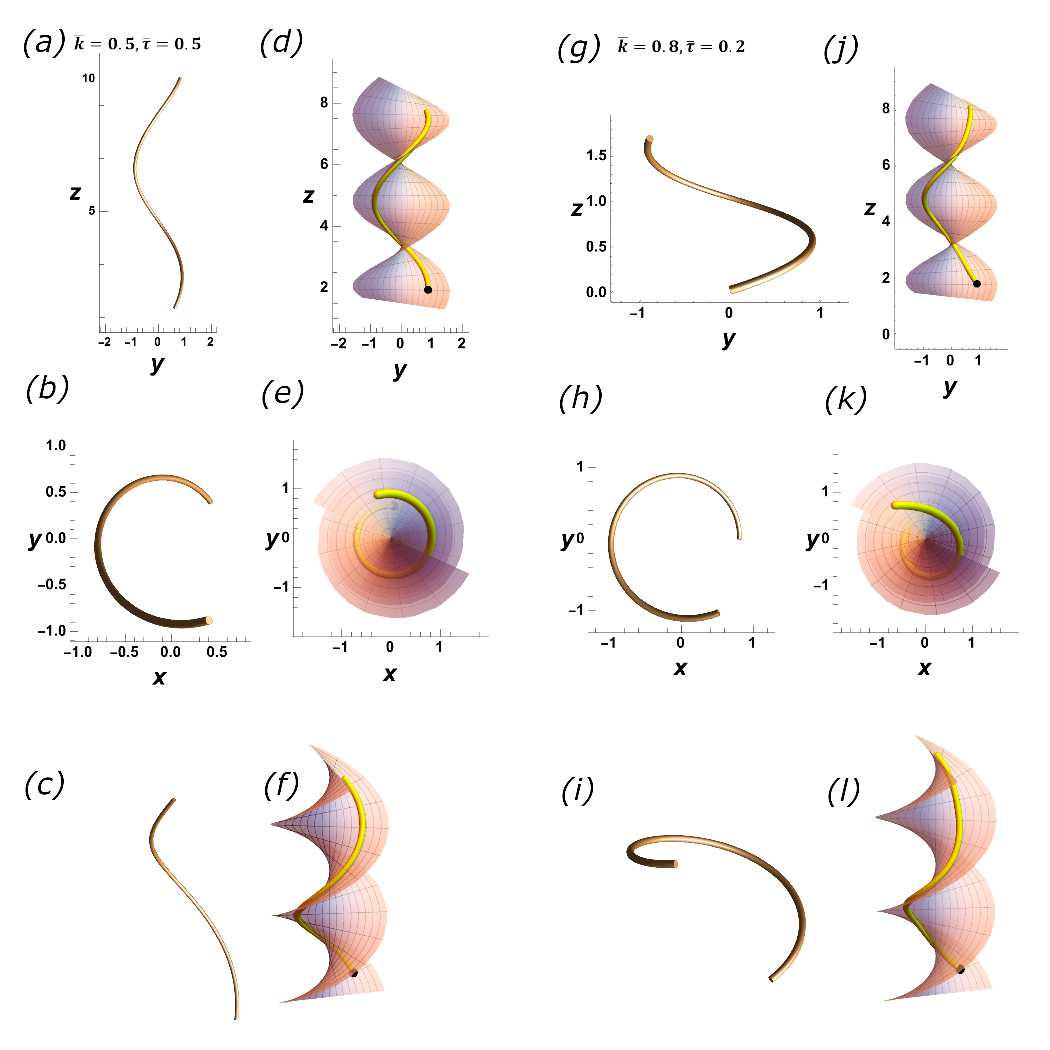}
        \caption{\textbf{Inextensible helices on a helicoid.} \textbf{(a)--(c)} A helix of normalized length $\zeta=3 \pi$ with  $\bar{k}=0.5$, $\bar{\tau}=0.5$ as the reference configuration. \textbf{(d)--(f)} Compatible state of the undistorted helix on a helicoid with matching pitch. \textbf{(g)--(i)} A helix of normalized length $\zeta=3 \pi$ with 
        $\bar{k}=0.8$, $\bar{\tau}=0.2$ as the reference configuration. \textbf{(j)--(l)} The current, distorted configuration of the helix on the same helicoid at equilibrium. From top to bottom: side, top, and elevated views. The black dots represent the tethered ends of the curves.}
        \label{fig:helicoid1}
    \end{figure*}   
        \begin{figure}[t]
        \centering
        \includegraphics[width=\columnwidth] {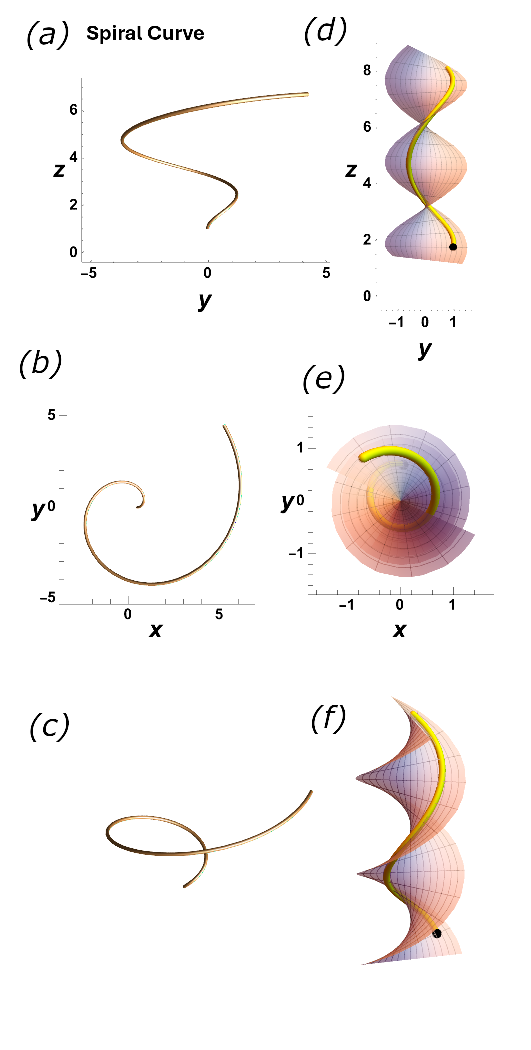}
        \caption{\textbf{Inextensible, incompatible spiral on a helicoid.} \textbf{(a)--(c)} A 3D spiral of normalized length $\zeta=2 \pi$ with the curvature and torsion given by Eq.~\ref{eq:spiral}.   \textbf{(d)--(f)} The finite energy state of the spiral on the helicoid. From top to bottom: side, plan, and elevated views. The black dots are the tethered points of the curve.}
        \label{fig:helicoid-var}
    \end{figure}   
    
    Helicoid is another example of a minimal surface and is parametrized by,
    \begin{align}
        \mathbf{X}(u,v) = \big( u\cos{v}, u\sin{v}, cv \big)\,, \label{eq.53}
    \end{align}
    where $u$ is the radial direction. For constant $u$, the periodic functions in Eq.~\ref{eq.53} cause the surface normal $\mathbf{\hat{N}}$ to twist as it moves in the vertical direction $v$. The pitch $c$ determines how fast $\mathbf{\hat{N}}$ twists; we set $c=1$. Per Eq.~\ref{eq.53}, the diagonal elements of the second fundamental form of the surface vanish ($b_{11}=b_{22}=0$). Consequently, a straight line can be at the ground state on the helicoid when attached to the central axis (Fig.~\ref{fig:helicoid-st}, yellow curve). In this configuration, $\tau_g=1$ such that the Darboux frame twists at a constant angle along the surface. If, instead, the tethering point is moved away from the central axis of the helicoid, the line orients itself in the radial direction, which is still a compatible state (Fig.~\ref{fig:helicoid-st}, red curve).
    
    Although a helicoid is composed of a family of helices located at $u:$ constant, those with a different pitch will not be compatible with a helicoid for a given value of $c\,.$ Fig.~\ref{fig:helicoid1} shows two helical curves that have constant curvature and torsion bounded on the helicoid surface. The first helix with $\bar{k}=0.5, \bar{\tau}=0.5$ is compatible with the surface when tethered at $u_0=1$. In this configuration, $k_N=k_{N,2}=0$ and $k_g=\tau_g=0.5$, and the space curve does not deform when confined on the surface because of its compatibility (Fig.~\ref{fig:helicoid1}a--f). The second helix with $\bar{k}=0.8, \bar{\tau}=0.2$, bounded at the same point on the surface, is not compatible since it twists slower than the background surface it is attached to (Fig.~\ref{fig:helicoid1}g--l). So, the curve deviates from a constant $u$ trajectory on the surface to reduce its geodesic torsion $\tau_g$ and increase its geodesic curvature to match its total curvature (as the normal curvature $k_N=0$).

    Fig.~\ref{fig:helicoid-var} demonstrates a spiral curve (conical helix) with the parametric equation
    \begin{align}
        \mathbf{X}_c(\sigma)= \big(\sigma \cos{\sigma}, \sigma \sin{\sigma}, \sigma \big)\,,
    \end{align}
    tethered to the helicoid. This leads to the variable intrinsic curvature and torsion,
    \begin{gather}
        \bar{k} = \sqrt{\frac{8+5\sigma^2+\sigma^4}{(2+\sigma^2)^3}}\,,\quad
        \bar{\tau} = \frac{(6+\sigma^2)}{(8+5\sigma^2+\sigma^4)}\,. 
        \label{eq:spiral}
    \end{gather}
    Both the intrinsic curvature and torsion of the spiral curve decreases as it moves upwards and opens up outwards. But on the helicoid surface, helical curves on the outer rims have higher geodesic curvature and lower geodesic torsion, while the normal curvature vanishes. So, the curve has to balance between matching the geodesic torsion deformation while not increasing the bending mismatch. The equilibrium configuration is highly stressed. For the spiral curve, we optimize the curve configuration using the downhill simplex method to find the tethered point corresponding to a local energy minimum.

    \subsection{Extensible curves}
    When a curve is highly incompatible on a surface with a locally changing metric tensor and the second fundamental form, geometric frustration can in principle distribute the energy cost between stretching and bending. To study this case, we consider the full formulation of our theory for extensible curves given by Eqs.~\ref{eq.23} and~\ref{eq.26}.  We show that the stretching term of an incompatible curve becomes important for two exemplary cases: first, for curves at global equilibrium on a cylinder, and second, for curves at local equilibrium on a hyperbolic paraboloid, where the coefficients of the metric and the second fundamental form tensors are functions of the surface coordinates $u,v\,.$ When we define axial strain for the elastic curve as 
    \begin{equation}
        \varepsilon_A = \frac{1}{2} \left(1-\frac{1}{\Lambda^2}\right)\,,
    \end{equation} 
    the unitless axial stress $\eta(\sigma)$ conjugate to the strain is found as
    \begin{gather}
        \eta(\sigma)\equiv \frac{w_{1D}^C}{\partial \varepsilon_A} = \frac{1}{2} \left(1-\frac{1}{\Lambda^2}\right)\,, \label{eq.48a}
    \end{gather}
    where $\eta(\sigma)$ is nondimensionalized by the Young's modulus $Y$. To visualize stretching, we will color code $\eta(\sigma)$ along the curves.
    
    \subsubsection{Cylinder}
    \label{sec:cylinder_inextensible}

    \begin{figure*}[ht]
        \centering
        \includegraphics[width=\textwidth]{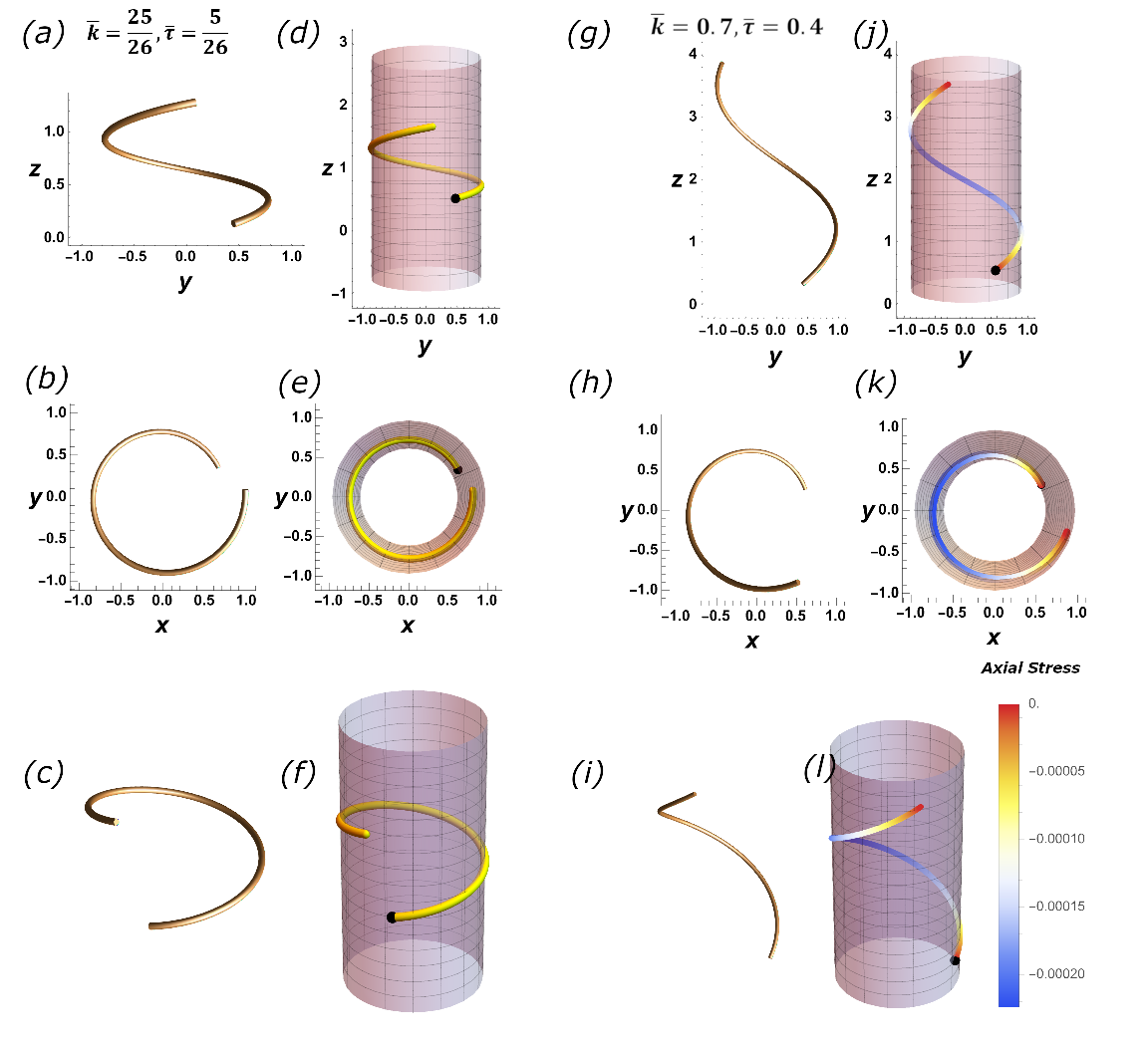}
        \caption{\textbf{Extensible curves on a cylinder.} \textbf{(a)--(c)} A helix with the normalized length $\zeta=2\pi$ with $\bar{k}=25/26$, $\bar{\tau}=5/26$ in its reference configuration. \textbf{(d)--(f)} The current compatible configuration of the helix on the cylinder. \textbf{(g)--(i)} A helix with the normalized length $\zeta=2\pi$ with with $\bar{k}=0.7$ , $\bar{\tau}=0.4$ in its reference configuration. Incompatible helical curve on the cylinder surface that has both bending and stretching deformation. \textbf{(j)--(l)} The current configuration of the helix with residual stress on the cylinder. The colormap indicates the unitless axial stress (Eq.~\ref{eq.48a}). The panels show the  side (a, d, g, j), plan (b, e, h, k), and elevated views (c, f, i, l). The black dots are the tethered points of the curves.}
        \label{fig:cylinder-helix}
    \end{figure*} 
    
    Planar curves are not the only possible geodesics on a unit cylinder introduced in Sec.~\ref{sec:cylinder_inextensible}. Circular helices with a rate of rotation $\sqrt{\bar{k}^2 + \bar{\tau}^2}=1$ would also form compatible geodesics on the cylinder with unit radius. However, as with the helicoid, if the curvature and torsion combination of the helix is incompatible with the cylinder, the curve will deform and become frustrated. In Fig.~\ref{fig:cylinder-helix}, we show a compatible and an incompatible helix on a cylinder for comparison. The compatible helix follows a geodesic path on the surface (Fig.~\ref{fig:cylinder-helix}a--f). For the incompatible helix with $\bar{k}=0.7\,,\,\bar{\tau}=0.4\,,$ the equilibrium configuration is a curve that locally bends further to fit the higher curvature of the surface and is compressed along its length to reduce the bending energy penalty. This example showcases an emergent stretching deformation when the curve is incompatible, albeit with a small axial stress on a constant curvature surface like the cylinder.

    \subsubsection{Hyperbolic paraboloid}

       \begin{figure}[t]
        \centering
        \includegraphics[width=\columnwidth]{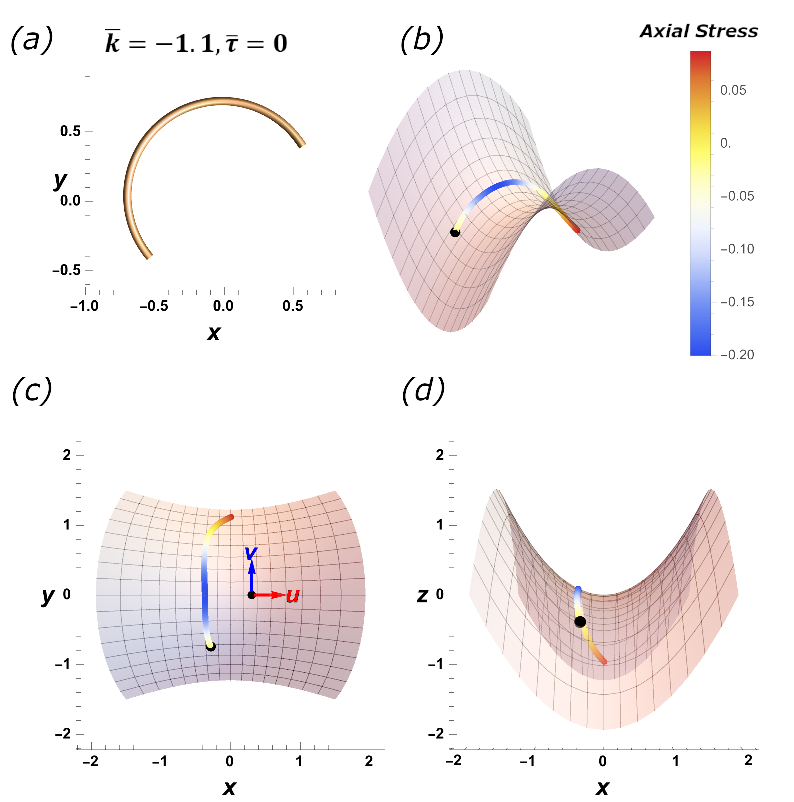}
        \caption{\textbf{Extensible curves on a hyperbolic paraboloid.} \textbf{(a)} A reference plane curve with the normalized length $\zeta=4$ and the reference curvature $\bar{k}=-1.1\,.$ \textbf{(b)--(d)} The elevated, top, and side views of the stressed current configuration that involves both bending and stretching. The colormap indicates the unitless axial stress (Eq.~\ref{eq.48a}). The black dots are the tethered points of the curves.}
        \label{fig:hyper-1.1}
    \end{figure}  
    \begin{figure}[t]
        \centering
        \includegraphics[scale=0.7]{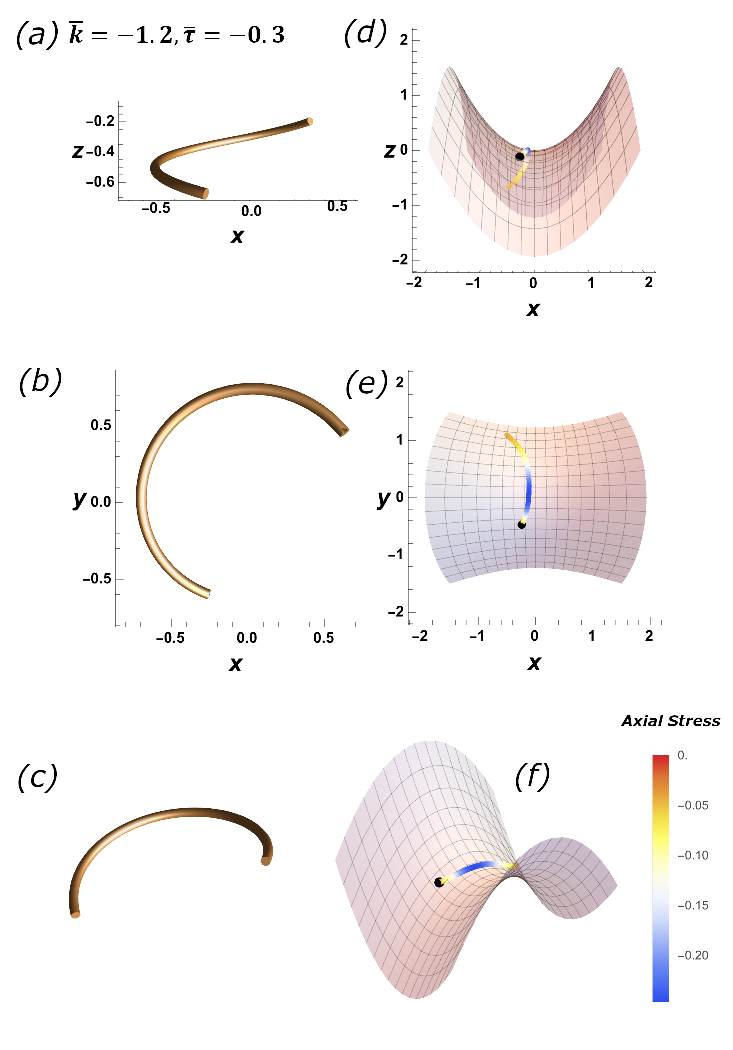}
        \caption{\textbf{Extensible curves on a hyperbolic paraboloid (continued).} \textbf{(a)--(c)} A reference helix with the normalized length $\zeta=3$ and the reference variables $\bar{k}=-1.2$, $\bar{\tau}=-0.3$. \textbf{(d)--(f)} The current configuration of the helix on the surface. The colormap indicates the unitless axial stress (Eq.~\ref{eq.48a}).
        From top to bottom: side view, top view, and elevated views. The black dots are the tethered points of the curves.}
        \label{fig:hyper-1.1-0.2}
    \end{figure}      

    Hyperbolic paraboloid is a saddle surface parametrized by,
    \begin{align}
        \mathbf{X}(u,v) = \left( u, v, \frac{1}{2}(u^2-v^2) \right)\,, \label{eq.58}
    \end{align}
    where $u$ is the concave, positive curvature direction and $v$ is the convex, negative curvature direction (Fig.~\ref{fig:hyper-1.1},~\ref{fig:hyper-1.1-0.2}). The metric components, the Gaussian curvature $K$, and the mean curvature $H$ of the surface are all functions of $u\,,\,v\,.$ Thus, any curve with either constant curvature or constant torsion is incompatible on the surface, and energy minimization gives rise to residual bending and stretching deformations. Here, we present the simulations for the equilibration of a planar reference curve and a helical reference curve on the hyperbolic paraboloid. In both cases, we optimize the configuration using the downhill simplex method to find the local minima, which yield residual axial stress.
    
    In Fig.~\ref{fig:hyper-1.1}, a planar curve with the normalized length $\zeta=4$ and negative reference curvature $\bar{k}<0$ is initially positioned along constant $u=0$ with a geodesic configuration. After optimization, the curve reaches its local minimum at the tethered point $(u_0,v_0) = (-0.4, -1.04)$ (Movie~S1)~\cite{kaplan_1D_elasticity}. The middle portion of the equilibrium curve slides toward the negative $u$ direction where it distributes the bending frustration among finite surface curvatures $k_g$ and $k_N$ and geodesic torsion $\tau_g$. As a result, the middle portion is highly compressed where the intrinsically bent curve can maintain lower bending energy cost at a region where surface curvature rapidly changes, while the two ends are slightly stretched. Likewise, the equilibrium configuration of the helix of the normalized length $\zeta=3$ in Fig.~\ref{fig:hyper-1.1-0.2} is shifted towards the positive $v$ side of the surface where again the intrinsic curvature and torsion of the reference curve are better accommodated at a cost of nonuniform compression. The optimal tethered point is found to be at $(u_0,v_0) = (-0.33, -0.62)$ (Movie~S2)~\cite{kaplan_1D_elasticity}. Altogether, Figs.~\ref{fig:hyper-1.1} and~\ref{fig:hyper-1.1-0.2} demonstrate how curves can accumulate stretching energy on complex, curved surfaces to minimize bending frustration.

\section{Discussion and conclusions}
\label{sec.6}
    We have developed an effective 1D theory of flexible slender objects modeled as elastic curves that can bend and stretch. Since our theory is coarse-grained from the surface theory of thin elastic shells, the bending terms arise in terms of the three Darboux curvatures of a surface-bound curve. Consequently, in relevance to various biological phenomena, our theory allows for a natural treatment of elastic curves confined on a background surface as showcased herein on five different curved surfaces, as opposed to previous works that focused on surfaces with high symmetries, such as cylinders and spheres. Our results for extensible curves on a cylinder and hyperbolic paraboloid surface motivate the need of a general theory that takes into account stretching deformations. 

     While our theoretical framework can mathematically predict the conformation of both compatible and incompatible curves on complex surfaces, our results remain to be validated by experiments of filaments confined to different surfaces. In this regard, the work by Prasath et al. on the shape of silicone-melt filaments on the spherical soap bubble surface could potentially be extended to explore other surface geometries~\cite{prasath2021shapes}.

     On the theoretical side, a natural generalization would be to relax the rigid surface assumption by considering soft elastic or fluid membranes than can deform under the stresses from the curve, closing the feedback loop between the mechanics of the curve and that of the surface. This would necessitate a coupling between the 1D theory of surface curve and a 2D theory of deformable membranes~\cite{sharma2023computational, shi2023morphological, zhang2026euler, vazquez2025equatorial}.  Mathematically, the energy functional would no longer just depend on the local variables along the curve, but also to the global surface variables that change through surface deformations induced by the curve. Furthermore, delamination of filaments off the membrane can also be investigated when the hard surface constraints are replaced by soft constraints by squaring the relevant terms in in Eq.~\ref{eq.26}. 
     
     Another potential direction is to investigate the statistical mechanics of surface curves by calculating the thermodynamic properties such as the orientational correlation functions and the curve persistence length altered by the surface confinement. It would also be valuable to assess how those measures compare with those of the corresponding reference space curve, or for a ribbon-like filament confined on the surface by changing the aspect ratios present in the theory~\cite{yong2022statistics,panyukov2000fluctuating, grossman2018shape, grossman2016elasticity}. 

     In conclusion, the theory presented in this work demonstrates a passive feedback mechanism due to the coupling between geometry and mechanics, which most likely works together with active biochemical feedback cycles underlying various biological phenomena. Elucidating this mechanism may thus contribute to the rational design of bioinspired matter that employs soft deformation pathways in applications ranging from synthetic biology to soft robotics.

\begin{acknowledgments}
We thank A. Concha, E. Cerda, S. Cheng, B. Greenvall, J. A. Hanna, S. Kale, C. Moguel-Lehmer, H. Robinson for fruitful discussions and the VT College of Science for financial support. We acknowledge the VT Advanced Research Computing Center for computing resources.
\end{acknowledgments}

\section*{Data Availability}
The supplementary movies S1 and S2 and the data and codes that support the findings of this article are openly available~\cite{kaplan_1D_elasticity}.

\appendix

\counterwithin*{equation}{section}
\renewcommand\theequation{\thesection\arabic{equation}}


    \section{Frenet-Serret equations for the reference space curve}\label{ap B}
    
    The configuration of the reference curve can be described by the Frenet-Serret material frame as defined in Sec.~\ref{sec.2}. The Frenet-Serret triad vectors ($\partial_{\bar{s}}\mathbf{\bar{R}}\perp \; \mathbf{\bar{L}} \perp \;\mathbf{\bar{B}}$) are used to define the reference curvature $\bar{k}$ and torsion $\bar{\tau}$ of the curve in space, such that they follow the Frenet-Serret equations given by
    
    \begin{equation}   
    \begin{bmatrix}
    \frac{\partial^2 \mathbf{\bar{R}}}{\partial \bar{s}^2} \\  \\
    \frac{\partial \mathbf{\bar{L}}}{\partial \bar{s}} \\ \\
    \frac{\partial \mathbf{\bar{B}}}{\partial \bar{s}}
    \label{eq:FrenetSerret}
    \end{bmatrix} 
    =
    \begin{bmatrix}
    0 && \bar{k} && 0 \\ \\
    -\bar{k} && 0 && \bar{\tau} \\ \\
    0 && -\bar{\tau} && 0
    \end{bmatrix}
    \begin{bmatrix}
    \frac{\partial \mathbf{\bar{R}}}{\partial \bar{s}} \\  \\
    \mathbf{\bar{L}} \\ \\
    \mathbf{\bar{B}} 
    \end{bmatrix} \,.
    \end{equation}



    \section{3D to 2D elasticity in Almansi strain}\label{ap D}

    The incompatible elastic energy of a 3D manifold, constituted with the invariants of Almansi strain has the form (the elasticity tensor $A^{ijkl}$ and strain tensor $\epsilon_{ij}$ defined in Eqs.~\ref{eq:elasticity_tensor_main}, \ref{eq:strain_tensor_main})
    \begin{equation}
        E_{3D} = \int d\bar{V} \frac{1}{2}  A^{ijkl} \epsilon_{ij} \epsilon_{kl}. \label{D.1}
    \end{equation}
    Taking the variation of this energy functional would give the Euler-Lagrange equations governing the equilibrium configuration of the elastic body.
    
    With the current metric as the measure (see Sec.\ref{sec.2} for the definitions), the following relations hold:
    \begin{gather}
        g^{ij} = (g^{-1})^{ij} \Rightarrow g^{ij}g_{ij} = \delta^i_i , \nonumber \\ 
        g^{ij}\bar{g}_{ij} = \bar{g}^i_i \neq \delta^i_i ,\nonumber \\
        \delta \bar{g}_{ij} = 0 ,\nonumber \\
        \delta \bar{g}^{ij} = \delta (g^{ii} g^{jj} \bar{g}_{ij}) \neq 0 ,\nonumber \\
        \delta g_{ij} = 2 \; \partial_i \mathbf{Y} \cdot\partial_j \mathbf{Y}, \nonumber \\
        \delta g^{ij} = -2 \; \partial^i \mathbf{Y} \cdot\partial^j \mathbf{Y}. \nonumber        
    \end{gather}
    So, the variation of the elastic energy functional,
    \begin{gather}
        \delta E_{3D} = \int d\bar{V} \frac{1}{2} \left[ \left( \delta A^{ijkl} \right) \epsilon_{ij} \epsilon_{kl} + A^{ijkl} \delta \left(  \epsilon_{ij} \epsilon_{kl} \right)  \right],  \nonumber
    \end{gather} 
     gives in terms of the 3D stress tensor $S^{ij}\equiv A^{ijkl} \epsilon_{kl} + \mathcal{O}[(g^{-1}\epsilon)^2]$ the 3D Euler-Lagrange equations and their boundary conditions as ($N_j:$ components of the surface normal vector)
    \begin{gather}
        \nabla_i \left( \sqrt{\frac{|\bar{g}|}{|g|}} S^{ij} \right) = 0\,, \label{D.2}\\ 
        S^{ij} N_j = 0\,. \label{D.3}    
    \end{gather}
    For bodies with thickness much smaller than the width and the length ($h\ll W,\, \ell$), we can effectively model them as 2D objects whose elastic behavior can be captured by the elastic behavior of the midsurface at $x^3=z=0$.     
    In the bulk of this thin body, first, we use the 3D boundary conditions Eq.~\ref{D.3} in the direction normal to the midsurface, or what constitutes the modified Kirchhoff-Love plane-stress assumption, and thereafter the plane-strain assumption,   
    \begin{gather}
        S^{i3}N_3=0, \nonumber \\
        \Rightarrow S^{i3} = \left( \lambda g^{i3} \epsilon_k^k + 2 \mu \epsilon^{i3} \right) = 0, \nonumber \\
        \Rightarrow \epsilon_3^3 = - \frac{\lambda}{\lambda+2\mu}\epsilon_\alpha^\alpha; \,\, \epsilon^3_{\alpha} = 0, \label{eq.3d_approx}    
    \end{gather}
    where the indices $\alpha, \beta \equiv \{1,2\} $ are defined on the midsurface. \ref{eq.3d_approx}  simplifies the 3D energy expression, as the strain tensor components normal to the midsurface can be written in terms of the midsurface components only
    \begin{gather}
        A^{ijkl}\epsilon_{ij}\epsilon_{kl} = \lambda \epsilon^i_i \epsilon^j_j + 2 \mu \epsilon^k_i \epsilon^i_k \nonumber \\ 
        = 2\mu \left( \frac{\lambda}{\lambda + 2\mu} \epsilon_\alpha^\alpha\epsilon_\beta^\beta + \epsilon_\beta^\alpha \epsilon_\alpha^\beta \right)\,, \nonumber \\
        = A^{\alpha \beta \gamma\lambda} \epsilon_{\alpha\beta} \epsilon_{\gamma\lambda}\,, \label{D.4}
    \end{gather}
    where
    \begin{gather}
        A^{\alpha \beta \gamma\delta} \equiv \mu \left[ \frac{2\lambda}{\lambda + 2\mu} g^{\alpha\beta} g^{\gamma\delta} + \left( g^{\alpha \gamma}g^{\beta\delta}  + g^{\alpha \delta}g^{\beta\gamma}  \right) \right]. \nonumber
    \end{gather}
    Next, we expand the 3D position vector about the midsurface in powers of $z$,\\
    \begin{gather}
        \mathbf{Y}(x^1, x^2, z) = \mathbf{Y}(x^1, x^2, 0) + z \frac{\partial \mathbf{Y}}{\partial z} \bigg|_{z=0} + \mathcal{O}(z^2), \nonumber \\
        = \mathbf{X}(x^1, x^2) + z \; \mathbf{\hat{N}} (x^1, x^2) + \mathcal{O}(z^2), \label{D.5}
    \end{gather}
    where,  $\mathbf{Y}|_{z=0}\equiv\mathbf{X}$ is the position vector of the midsurface, and $\frac{\partial \mathbf{Y}}{\partial x^{3}}\big|_{z=0} \equiv\mathbf{\mathbf{\hat{N}}}$ is the vector normal to the midsurface. Taking the derivatives of Eq.~\ref{D.5} with respect to the midsurface coordinates provides the following definitions of $a_{\alpha \beta}, b_{\alpha \beta}, c_{\alpha \beta}$ -- respectively, the first, second and third fundamental forms of the midsurface,
    \begin{gather}
        a_{\alpha\beta} \equiv \partial_{\alpha} \mathbf{X} \cdot \partial_{\beta} \mathbf{X}\,, \nonumber \\
        b_{\alpha\beta} \equiv \partial_{\alpha \beta} \mathbf{X} \cdot \mathbf{\mathbf{\hat{N}}}\,, \nonumber \\
        c_{\alpha\beta} \equiv \partial_{\alpha}\mathbf{\hat{N}}\cdot \partial_{\alpha}\mathbf{\hat{N}}\,, \nonumber 
    \end{gather}    
    and trivially, $g_{33} = 1 + \mathcal{O}(z^2),\; g_{\alpha 3}=0 + \mathcal{O}(z^2)$. We can also define the reference fundamental forms of the midsurface as $\bar{a}_{\alpha \beta}, \bar{b}_{\alpha \beta}, \bar{c}_{\alpha \beta}$. Consequently, the 2D strain tensor
    is
    \begin{gather}
    \varepsilon_{\alpha \beta} \equiv \frac{1}{2}(a_{\alpha \beta} - \bar{a}_{\alpha \beta}). \nonumber
    \end{gather}
    $\mathcal{A}^{\alpha \beta \gamma \delta}$ is the reduced 2D elasticity tensor in terms of the components of the current metric of the midsurface, and the Lam\'e constants $\lambda, \mu$ replaced by the canonical material parameters of the slender object, the Young's modulus $Y$ and the Poisson's ratio  $\nu$, such that
    \begin{gather}
        \mathcal{A}^{\alpha \beta \gamma \delta} = \frac{Y}{8(1-\nu^2)} \left( \nu \; a^{\alpha \beta} \; a^{\gamma \delta} + (1-\nu) \; a^{\alpha \gamma} \; a^{\beta \delta}\right), \nonumber 
        \\2\mu = \frac{Y}{1+\nu}\,, \quad \frac{\lambda}{2\mu+\lambda} = \frac{\nu}{1-\nu}\,.\nonumber
    \end{gather} 
    Then the factors in Eq.~\ref{D.4} have the series expansion,
    \begin{gather}
        A^{\alpha \beta \gamma \delta} = \mathcal{A}^{\alpha \beta \gamma \delta} + 2z \left( \mathcal{A}^{\lambda \beta \gamma \delta} b_\lambda^\alpha + \mathcal{A}^{\lambda \beta \lambda \delta} b_\lambda^\gamma \right) + \mathcal{O}(z^2) \label{D.6}, \\
        \epsilon_{\alpha \beta} = \varepsilon_{\alpha \beta} - z \;(b_{\alpha \beta} - \bar{b}_{\alpha \beta}) + \frac{z^2}{2} (c_{\alpha \beta} - \bar{c}_{\alpha \beta}) + \mathcal{O}(z^3). \label{D.7}  
    \end{gather}
    Finding the explicit $z$-dependence of the energy here allows us to complete the first coarse-graining step. Substituting Eqs. \ref{D.6} and \ref{D.7} to Eq.~\ref{D.4} and integrating over the thin dimension $z \in [h/2, -h/2]$ in Eq.~\ref{D.1}, we obtain the reduced 2D energy functional
    \begin{gather}
    E_{2D} = \int dA \sqrt{\frac{|\bar{a}|}{|a|}} \mathcal{A}^{\alpha \beta \gamma \delta} \bigg( \frac{h}{2}
    \varepsilon_{\alpha \beta}\;\varepsilon_{\gamma \delta}  \nonumber \\
    + \frac{h^3}{24}
    (b_{\alpha \beta}-\bar{b}_{\alpha \beta})\;(b_{\gamma \delta}-\bar{b}_{\gamma \delta}) + \mathcal{O}(b^2\delta, b^2\delta^2) \bigg). \label{D.8}
    \end{gather}
    Due to the the small stretch approximation: $a^{11} \varepsilon_{11} \equiv \delta \ll 1$ in the 1D theory, it can be shown that the other mixed $h^3$ terms in the bending content give higher order terms in the Euler-Lagrange equations~\cite{hanna2019some}. We have also made the approximation that $\sqrt{\frac{|\bar{g}|}{|g|}} \simeq \sqrt{\frac{|\bar{a}|}{|a|}}$, neglecting higher order contributions to the bending energy. So, we only keep the terms quadratic in $\varepsilon, b$.


    Eq.~\ref{D.8} describes the elasic behavior of a thin plate ($\bar{b}_{\alpha\beta},  \bar{c}_{\alpha\beta} = 0$) or a shell ($\bar{b}_{\alpha\beta},  \bar{c}_{\alpha\beta} \neq 0$). To find the equilibrium surface configuration, we vary the energy functional with respect to $a_{\alpha \beta}$ and  $b_{\alpha \beta}$, expressed in terms of the position vector and its derivatives. Then, the 2D moment and augmented stress tensors are defined as, 
    \begin{gather}
     \mathcal{L}^{\mu \nu} =  \sqrt{\frac{|\bar{a}|}{|a|}} M^{\mu \nu} = \sqrt{\frac{|\bar{a}|}{|a|}} \frac{\partial w_{2D}}{\partial b_{\mu\nu}}, \label{D.9} \\   
     \mathcal{K}^{\mu \nu} =  \sqrt{\frac{|\bar{a}|}{|a|}} (S^{\mu \nu} - 2 Q^{\mu\nu}) = \sqrt{\frac{|\bar{a}|}{|a|}} \left( \frac{\partial w_{2D}}{\partial \varepsilon_{\mu\nu}} + \frac{\partial w_{2D}}{\partial a_{\mu\nu}} \right),  \label{D.10} 
     \end{gather}
    where the extra second term in $\mathcal{K}^{\mu\nu}$ comes from the explicit current metric components in the elasticity tensor. The Euler-Lagrange equations for the moment balance and stress balance in the bulk are respectively given as 
    \begin{gather}
        \nabla_{\mu \nu} \mathcal{L}^{\mu \nu}  - \mathcal{L}^{\mu \nu} c_{\mu \nu} - \mathcal{K}^{\mu \nu}b_{\mu \nu} = 0\,, \label{D.11} \\
        \nabla_\nu\left( \mathcal{K}^{\mu\nu} +\mathcal{L}^{\gamma \nu}b_{\nu}^{\mu} \right) + (\nabla_\gamma \mathcal{L}^{\gamma \nu})b^\mu_\nu = 0\,, \label{D.12} 
    \end{gather}
    with the natural boundary conditions ($\hat{n}:$ unit vector pointing orthogonal to the edges of the sheet),
    \begin{gather}
        (\nabla_\mu \mathcal{L}^{\mu \nu} )n_{\nu} = 0\,, \label{D.13} \\
         \mathcal{L}^{\mu \nu} n_{\mu} n_{\nu}  = 0\,, \label{D.14} \\
         \left( \mathcal{K}^{\mu\nu} +\mathcal{L}^{\gamma \nu}b_{\gamma}^{\mu} \right)  n_{\nu} = 0\,,  \label{D.15}
    \end{gather}

    \section{Other constitutive models} \label{ap E}
    Our consitutive model expresses the elasticity tensor in terms of the current metric, as given in Eq.~\ref{eq:elasticity_tensor_main}. Instead, if we had followed the formulation of Efrati et al.~\cite{efrati2009elastic} where the elasticity tensor components are defined in terms of the reference metric as
    \begin{gather}
        \bar{A}^{ijkl} \equiv  \left[ \lambda \; \bar{g}^{ij} \bar{g}^{kl} + \mu \; \left( \bar{g}^{ik} \bar{g}^{jl} + \bar{g}^{il} \bar{g}^{jk} \right) \right],
    \end{gather}
    the energy functional in Eq.~\ref{eq.34} would then become
    \begin{widetext}
    \begin{equation}
        E_{1D} = \frac{YhW\ell}{8} \int \sqrt{\bar{g}} \; d\sigma \left\{ \left( \Lambda^2-1 \right)^2 + \frac{\gamma^2}{3} \left[ \Lambda^4 ( k_g - \frac{1}{\Lambda^2} \bar{k}_g )^2 \right.\right.
        \left.\left. + \Lambda^4 ( k_N - \frac{1}{\Lambda^2} \bar{k}_N )^2 + \frac{2 \Lambda^2}{1+\nu} ( \tau_g - \frac{1}{\Lambda} \bar{\tau}_g)^2 \right]  \right\} + \mathcal{O}(\gamma^2 \delta, \gamma^2 \delta^2)\,. \label{E.1}
    \end{equation}
    \end{widetext}
    This generates two different physical outcomes from Eq.~\ref{eq.23a}. First, the stretching term -- now in Green strain -- would diverge upon elongation but be finite for compression to $\Lambda=\sqrt{\frac{g}{\bar{g}}}\to 0\,.$ Second, the apparent coupling between stretching and bending manifested by the prefactors in powers of $\Lambda$ in the bending terms leads to the following behavior: If a circularly arc is stretched without changing the radius of curvature, the bending energy increases. That is, extending a bent rod would cost both bending and stretching energy. In contrast, Eq.~\ref{eq.23a} decouples stretching and bending deformations, which is better suited for our purposes.

    \section{Derivation of mapping equations} \label{ap F}
    In this section, we derive expressions for the curve variables in terms of the surface parameters $u,v$ to map the position of the curve on the surface. Let $\mathbf{R}(s)$ with the arc-length parameter $s$, be an embedded curve on the surface $\mathbf{X}(u,v)$ as defined in Sec.~\ref{sec.2}, such that, 
    \begin{align}
        s \mapsto \mathbf{R}(s) = \mathbf{X}(u(s),v(s))|_c.
    \end{align} 
    Here, $\mathbf{X}|_c$ is the position vector on the surface only along the curve. Consequently, $g_{\alpha\beta}|_c = a_{\alpha\beta}$ and $L_{\alpha\beta}|_c=b_{\alpha\beta}$.
    Using primes to denote the derivatives with respect to $s$, the tangent vector to the curve is given by
    \begin{gather}
        \frac{\partial \mathbf{R}}{\partial s} = \mathbf{R}'(s)= (\partial_u \mathbf{X})|_c \;u' + (\partial_v \mathbf{X})|_c \;v',\label{F.2}
    \end{gather}
    and the curvature vector,
    \begin{gather}
        \frac{\partial^2 \mathbf{R}}{\partial s^2} = \mathbf{R}''(s)= (\partial_{uu} \mathbf{X})|_c \;u'^2 + (\partial_{u} \mathbf{X})|_c \;u'',  \nonumber \\
        + 2\;(\partial_{uv} \mathbf{X})|_c \;u' \;v' 
        + (\partial_{vv} \mathbf{X})|_c \;v'^2 + (\partial_{v} \mathbf{X})|_c \;v''. \label{F.3}
    \end{gather}
    If we now define a second direction $r$ on the surface that is normal to $s$ along the curve, we have an orthonormal triad of Darboux vectors for the curve constrained to the surface: curve tangent $\mathbf{R}'$, surface normal $\mathbf{\mathbf{\hat{N}}}$, and normal to the curve on the surface
    \begin{gather}
        \frac{\partial \mathbf{R}}{\partial r}  = \mathbf{\hat{n}} = \mathbf{\hat{N}}\times\mathbf{R}' \nonumber \\
        =\frac{1}{|a|} \bigg( u'\;(a_{uu} + v'\;a_{uv})(\partial_v\mathbf{R}) - (u'\;a_{uv} + v'\;a_{vv})(\partial_u\mathbf{R}) \bigg). \label{F.4}
    \end{gather}
    Eq.~\ref{F.4} allows us to define
    \begin{gather}
        \frac{\partial}{\partial r} \equiv \frac{1}{|a|} \bigg( (u'\;a_{uu} + v'\;a_{uv}) \frac{\partial}{\partial v} - (u'\;a_{uv} + v'\;a_{vv}) \frac{\partial}{\partial u} \bigg),
    \end{gather}
    so that,
    \begin{gather}
        \frac{\partial^2 \mathbf{R}}{\partial r^2} = \frac{1}{|a|}\bigg( (u'\;a_{uu} + v'\;a_{uv}) \frac{\partial}{\partial v} - (u'\;a_{uv} + v'\;a_{vv}) \frac{\partial}{\partial u} \bigg) \nonumber \\
        \bigg[ \frac{1}{|a|}\bigg( (u'\;a_{uu} + v'\;a_{uv}) \frac{\partial}{\partial v} - (u'\;a_{uv} + v'\;a_{vv}) \frac{\partial}{\partial u} \bigg) \bigg] \mathbf{R}. \label{F.6}
    \end{gather}
    Therefore, for the curve on the surface, we have the following identities,
    \begin{gather}
        1 = \frac{\partial \mathbf{R}}{\partial s}\cdot\frac{\partial \mathbf{R}}{\partial s}, \label{F.7} \\
        k_g(s) = \frac{\partial^2 \mathbf{R}}{\partial s^2}\cdot\mathbf{\hat{n}}, \label{F.8}\\
        k_N(s) = \frac{\partial^2 \mathbf{R}}{\partial s^2}\cdot\mathbf{\hat{N}}, \label{F.9}\\
        \tau_g(s) = \frac{\partial \mathbf{\hat{n}}}{\partial s}\cdot\mathbf{\hat{N}}, \label{F.10}\\
        k_{N,2}(s) =  \frac{\partial^2 \mathbf{R}}{\partial r^2} \cdot\mathbf{\hat{N}}. \label{F.11}
    \end{gather}
    Using the definition of the curve metric $g\equiv\frac{\partial \mathbf{R}}{\partial \sigma}\cdot\frac{\partial \mathbf{R}}{\partial \sigma},$ the substitution of Eqs.~\ref{F.2},~\ref{F.4},~\ref{F.6} to Eqs.~\ref{F.7}--\ref{F.11} leads to the surface mapping equations given by Eqs.~\ref{eq.44}--\ref{eq.48}.
 

\bibliography{ref2s}
\bibliographystyle{unsrt}  

\end{document}